\documentclass[12pt]{article}
\usepackage{amsbsy,amsfonts,amsmath,amssymb,amsthm,enumerate,framed}
\usepackage{color,graphicx}
\usepackage{rsfs}
\usepackage{bbm}
\newcommand{\ind}[1]{\mathbbm{1}_{\{#1\}}}
\allowdisplaybreaks[1]

\numberwithin{equation}{section}
\theoremstyle{plain}

\newtheorem{thm}{Theorem}[section]

\newtheorem{lmm}[thm]{Lemma}
\newtheorem{prp}[thm]{Proposition}
\newtheorem{rmk}[thm]{Remark}

\newcommand{\cD}{\mathcal{D}}
\newcommand{\cG}{\mathcal{G}}

\newcommand{\db}{\Longleftrightarrow}

\newcommand{\dpst}{\displaystyle}
\newcommand{\Exp}[1]{\langle #1 \rangle}
\newcommand{\EXP}[1]{\langle\!\langle #1 \rangle\!\rangle}
\newcommand{\lbeq}[1]{\label{eq:#1}}
\newcommand{\mB}{{\mathbb B}}

\newcommand{\mN}{\mathbb{N}}

\newcommand{\mR}{\mathbb{R}}
\newcommand{\mZ}{\mathbb{Z}}
\newcommand{\muc}{\mu_\text{c}}
\newcommand{\muNG}{\mu_N^{\sss(0)}}
\newcommand{\nn}{\nonumber}

\newcommand{\Proof}[1]{\medskip\noindent\textit{#1}~}
\newcommand{\QED}{\hspace*{\fill}\rule{7pt}{7pt}\medskip}

\newcommand{\refeq}[1]{(\ref{eq:#1})}
\newcommand{\sH}{\mathscr{H}}

\newcommand{\sJ}{\mathscr{J}}

\newcommand{\sss}{\scriptscriptstyle}
\newcommand{\sst}{\scriptstyle}
\newcommand{\tK}{\tilde K}
\newcommand{\tlamb}{\tilde\lambda}

\newcommand{\veee}[1]{|\!|\!|#1|\!|\!|}
\newcommand{\vep}{\varepsilon}
\newcommand{\vno}{\varnothing}
\newcommand{\vphi}{\varphi}
\newcommand{\vtri}{\vartriangle}

\newcommand{\wsigma}{\widetilde\sigma}
\newcommand{\wsigmavec}{\widetilde{\boldsymbol\sigma}}
\newcommand{\Zd}{\mZ^d}
\newcommand{\Zp}{\mZ_+}

\newcommand{\nvec}{\boldsymbol{n}}

\newcommand{\sigmavec}{\boldsymbol\sigma}

\newcommand{\vphivec}{\boldsymbol\vphi}
\newcommand{\olzd}{\,\mathop{\raisebox{-8pt}{\includegraphics[scale=0.2]
 {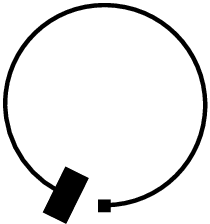}}}_{\tilde o}\,}

\title{Application of the lace expansion to the $\vphi^4$ model}
\author{
Akira~Sakai\thanks{Department of Mathematics, Hokkaido University, Japan.
\texttt{sakai@math.sci.hokudai.ac.jp}.}
}
\date{December 29, 2014}

\begin{document}
\maketitle

\begin{abstract}
Using the Griffiths-Simon construction of the $\vphi^4$ model and the lace
expansion for the Ising model, we prove that, if the strength $\lambda\ge0$
of nonlinearity is sufficiently small for a large class of short-range models
in dimensions $d>4$, then the critical $\vphi^4$ two-point function
$\Exp{\vphi_o\vphi_x}_{\muc}$ is asymptotically $|x|^{2-d}$ times a
model-dependent constant, and the critical point is estimated as
$\muc=\hat\sJ-\frac\lambda2\Exp{\vphi_o^2}_{\muc}+O(\lambda^2)$, where
$\hat\sJ$ is the massless point for the Gaussian model.
\end{abstract}

\tableofcontents

\section{Introduction and the main results}
The (lattice) $\vphi^4$ model is a pedagogical yet nontrivial model in scaler
field theory.  It is also considered to be an interface model defined by a
Hamiltonian having a quartic self-energy term.  (See, e.g., \cite{cd12} for
recent development in another class of interface models, called gradient
fields.)  If that quartic term is absent, then it becomes the Gaussian model
and its two-point function satisfies the same convolution equation as the
random-walk's Green function.  In particular, for the massless Gaussian model,
which is a lattice version of Gaussian free fields, the two-point function
decays as a multiple of $|x|^{2-d}$ as $|x|\to\infty$ when $d>2$.

On the other hand, the $\vphi^4$ two-point function is known to satisfy a
nonlinear equation, called the Schwinger-Dyson equation.  The nonlinearity is
due to involvement of four-spin expectations.  This implies that, in order to
find the exact expression for the two-point function, we must also know the
exact expressions for four-spin expectations.  In general, the Schwinger-Dyson
equation for $2n$-spin expectations involves $(2n+2)$-spin expectations.
Therefore, it is seemingly impossible to solve those infinitely many
simultaneous equations to find the exact expression for the two-point function.

Instead of solving those simultaneous equations, there have been many useful
ideas to study the phase transition and critical behavior for the $\vphi^4$
model.  Among those are the use of reflection positivity \cite{fils78,fss76}
and correlation inequalities obtained by the random-current representation
\cite{a82} and the random-walk representation \cite{bfs82,bfs83}.  They imply
that, for the nearest-neighbor model in dimensions $d>2$, there is a
nontrivial critical point $\muc\in\mR$ such that the two-point function
$\Exp{\vphi_o\vphi_x}_\mu$ is bounded above by a multiple of $|x|^{2-d}$
uniformly in $\mu>\muc$, and therefore all critical exponents take on their
mean-field values in dimensions $d>4$ \cite{a82,f82,s82} (see also
\cite{ffs92} and
references therein).  Moreover, for the nearest-neighbor model, the rigorous
renormalization-group (RG) approach based on the block-spin transformation
\cite{gk80,gk83,gk85} may identify an asymptotic expression for the critical
two-point function $\Exp{\vphi_o\vphi_x}_{\muc}$, which is presumably
$C|x|^{2-d}$ as $|x|\to\infty$ for some constant $C\in(0,\infty)$.  This is
proven to be affirmative when $d=4$ (cf., \cite[Theorem~I.2]{fmrs87} and
\cite[(8.32)]{gk84}; see also \cite{bbs14} for the recent RG results on the
$n$-component $|\vphi|^4$ model in 4 dimensions).  However, as far as we are
aware, such strong results have not been reported in dimensions $d>4$.

For the Ising model, which is considered to be in the same universality class
as the $\vphi^4$ model, we have been able to show \cite{s07} that, not only
for the nearest-neighbor model but also for a large class of spread-out models
which do not necessarily satisfy reflection positivity, the critical Ising
two-point function is asymptotically a model-dependent multiple of
$|x|^{2-d}$, if the dimension $d$ or the range of spin-spin coupling is
sufficiently large.  The proof is based on the lace expansion, which was first
applied to weakly self-avoiding walk \cite{bs85} and then developed for
lattice trees and lattice animals \cite{hs90}, percolation \cite{hs90'},
oriented percolation \cite{ny93} and the contact process \cite{s01}.  The
asymptotic behavior of the critical two-point function for each spacial model
is proved in \cite{h08,hhs03,s07}.  The methodology has been extended to
cover the case of power-law decaying spin-spin coupling \cite{cs14} (see
also \cite{hhs08} for results in the Fourier space).

In this paper, we apply the lace expansion for the Ising model to prove
asymptotic behavior of the $\vphi^4$ two-point function.  In order to do so,
we first use the Griffiths-Simon construction \cite{sg73} to approximate each
$\vphi^4$ spin by a sum of $N$ Ising-spin variables.  This is a well-known
approach to study the $\vphi^4$ model (see, e.g., \cite[Section~10]{a82}).
Then, we investigate the lace-expansion coefficients and determine the right
scaling in powers of $N$.  As a result, we prove the expected asymptotic
behavior of the critical two-point function, i.e.,
$\Exp{\vphi_o\vphi_x}_{\muc}\sim\exists C|x|^{2-d}$ as $|x|\to\infty$, for
a large class of short-range models on $\Zd$ with $d>4$, if the strength
$\lambda\ge0$ of nonlinearity is sufficiently small.  This implies
triviality of the continuum limit, as pointed in \cite[Section~7]{f82}
(see also \cite{a82}).  During the course, we also
obtain the $\lambda$-expansion of the critical point $\muc$ up to
$O(\lambda^2)$ around the massless point for the Gaussian model.

Before showing the precise statement of the above result, we first provide
the precise definition of the model.

\subsection{The $\vphi^4$ model}
For a finite set $\Lambda\subset\Zd$, we define the Hamiltonian $\sH_\Lambda$
on the space $\mR^\Lambda$ of spin configurations as follows: for
$\vphivec\equiv(\vphi_x)_{x\in\Lambda}\in\mR^\Lambda$,
\begin{align}\lbeq{Hamiltonian}
\sH_\Lambda(\vphivec)=-\frac12\sum_{u\ne v\in\Lambda}\sJ_{u,v}\vphi_u\vphi_v
 +\sum_{v\in\Lambda}\bigg(\frac\mu2\vphi_v^2+\frac\lambda{4!}\vphi_v^4\bigg),
\end{align}
where $\sJ_{u,v}$ is a nonnegative, translation-invariant and $\Zd$-symmetric
coupling function: $\sJ_{u,v}=\sJ_{o,v-u}\equiv\sJ(v-u)\ge0$.  We also assume
$\sJ(o)=0$ and that $\sJ$ is summable:
\begin{align}
\hat\sJ\equiv\sum_{v\in\Zd}\sJ(v)<\infty.
\end{align}
The parameter $\mu\in\mR$ plays the role of temperature, while $\lambda\ge0$
is the intensity of nonlinearity.  We call the model Gaussian if $\lambda=0$,
and in that case, we can rewrite the Hamiltonian as
\begin{align}\lbeq{GaussianModel}
\sH_\Lambda^{\sss\lambda=0}(\vphivec)=\frac{\hat\sJ}2(\vphivec,-\Delta\vphivec)
 +\frac{\mu-\hat\sJ}2|\vphivec|^2,
\end{align}
where $(\cdot,\cdot)$ is the inner product and $\Delta$ is the lattice
Laplacian defined by the transition probability $\sJ(x)/\hat\sJ$.  The first
term on the right-hand side represents the kinetic energy, while the second
term represents the potential.  We call the Gaussian model massless if the
potential is zero (i.e., $\mu=\hat\sJ$).

\begin{figure}[t]
\[ 0\hskip5pc\raisebox{1.5pc}{$\mu=\muc(\lambda)$}
 \hskip-5pt\raisebox{9pc}{$\lambda$}
 \hskip-5pt\raisebox{15pc}{$\chi_\mu$}
 \hskip1.3pc\raisebox{1.5pc}{$\mu=\hat\sJ$}
 \hskip-13.8pc\includegraphics[scale=0.6]{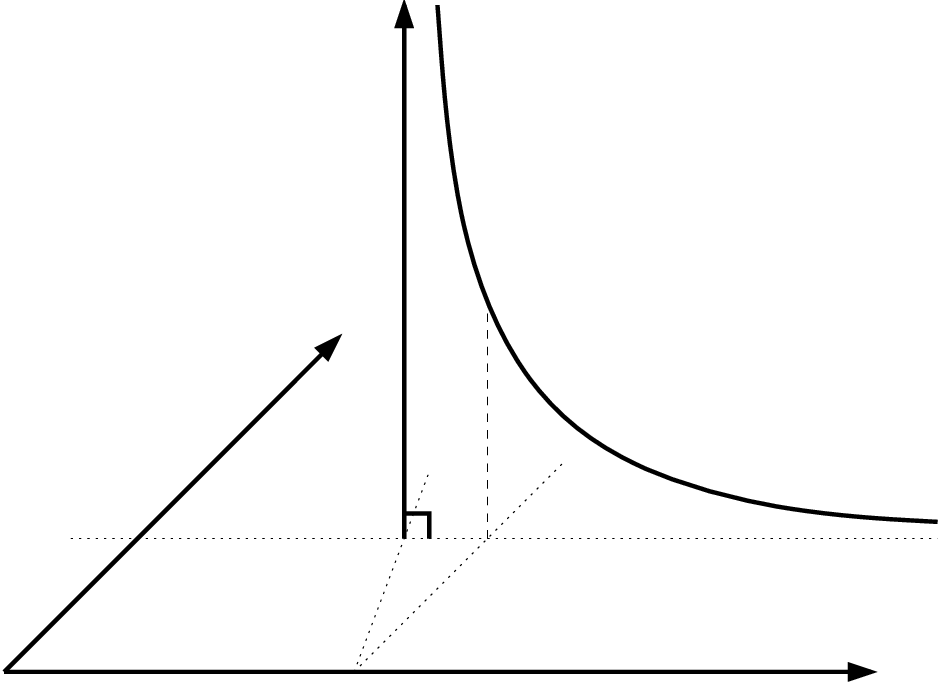}
 \hskip-1pc\mu \]
\caption{\label{fig:chi}Divergence of the susceptibility $\chi_\mu$ as
$\mu\downarrow\muc(\lambda)$ for a fixed $\lambda>0$ (along the horizontal
dotted line).  The susceptibility is
still finite below the massless line $\mu=\hat\sJ$, whereas the potential
$\sH_\Lambda(\vphivec)-\frac{\hat\sJ}2(\vphivec,-\Delta\vphivec)$ is not
convex any more.}
\end{figure}

The key quantity we are interested in is the two-point function
$\Exp{\vphi_o\vphi_x}_\mu$, which is the increasing limit as
$\Lambda\uparrow\Zd$ (due to the second Griffiths inequality, e.g.,
\cite{ffs92}) of the finite-volume expectation
$\Exp{\vphi_o\vphi_x}_{\mu,\Lambda}$:
\begin{align}
\Exp{\vphi_o\vphi_x}_\mu=\lim_{\Lambda\uparrow\Zd}\Exp{\vphi_o\vphi_x}_{\mu,
 \Lambda}\equiv\lim_{\Lambda\uparrow\Zd}\frac{\dpst\int_{\mR^\Lambda}\vphi_o
 \vphi_x\,e^{-\sH_\Lambda(\vphivec)}\,\text{d}\vphivec}{\dpst\int_{\mR^\Lambda}
 e^{-\sH_\Lambda(\vphivec)}\,\text{d}\vphivec}.
\end{align}
Due to Lebowitz' inequality \cite{l74}, there exists a critical point
$\muc\equiv\muc(d,\sJ,\lambda)\le\hat\sJ$ such that the susceptibility
\begin{align}
\chi_\mu\equiv\sum_{x\in\Zd}\Exp{\vphi_o\vphi_x}_\mu
\end{align}
is finite if and only if $\mu>\muc$ and diverges as $\mu\downarrow\muc$
(cf., Figure~\ref{fig:chi}).

One of the possible approaches to investigate the two-point function is
to use the result of integration by parts: $\Exp{\frac{\partial\sH_\Lambda}
{\partial\vphi_o}\vphi_x}_\Lambda=\delta_{o,x}$.  Plugging \refeq{Hamiltonian}
into this identity and taking the infinite-volume limit, we obtain the
Schwinger-Dyson equation
\begin{align}\lbeq{schwinger-dyson}
-\sum_v\sJ_{o,v}\Exp{\vphi_v\vphi_x}_\mu+\mu\Exp{\vphi_o\vphi_x}_\mu
 +\frac\lambda{3!}\Exp{\vphi_o^3\vphi_x}_\mu=\delta_{o,x}.
\end{align}
This immediately implies that the two-point function for the Gaussian model
satisfies the same convolution equation as the random-walk's Green function
generated by the 1-step distribution $\sJ/\hat\sJ$ with killing rate
$1-\hat\sJ/\mu$.  Therefore, at the massless point $\mu=\hat\sJ$, the
two-point function exhibits the same asymptotic behavior as the Green function
for the underlying random walk.  In this paper, in addition to the properties
stated below \refeq{Hamiltonian}, we assume that all moments of $\sJ$ are
finite.  In particular, we define the variance
\begin{align}\lbeq{Vdef}
V=\sum_{x\in\Zd}|x|^2\frac{\sJ(x)}{\hat\sJ}.
\end{align}
Then, by the Cram\'er-Edgeworth expansion (e.g., \cite[Theorem~A.1]{cs14}), we
can estimate the $n$-fold convolution of $\sJ/\hat\sJ$ and its ``derivative"
as (cf., \cite[(1.21) and (1.24)]{cs14})
\begin{align}
\frac{\sJ^{*n}(x)}{\hat\sJ^n}&\le\frac{O(n)}{(|x|\vee1)^{d+2}},\lbeq{sJ*n1}\\
\bigg|\frac{\sJ^{*n}(x)}{\hat\sJ^n}-\frac{\sJ^{*n}(x+y)+\sJ^{*n}(x-y)}{2\hat
 \sJ^n}\bigg|&\le\frac{O(n)|y|^2}{(|x|\vee1)^{d+4}}\qquad[|y|\le\tfrac13|x|].
 \lbeq{sJ*n2}
\end{align}
Consequently, by the standard random-walk analysis, we can readily show that
the massless Gaussian two-point function
$\Exp{\vphi_o\vphi_x}_{\hat\sJ}^{\sss\lambda=0}$ exhibits the asymptotic
behavior
\begin{align}
\Exp{\vphi_o\vphi_x}_{\hat\sJ}^{\sss\lambda=0}\underset{|x|\to\infty}\sim
 \frac{\frac{d}2\Gamma(\frac{d-2}2)\pi^{-d/2}}{\hat\sJ V|x|^{d-2}}.
\end{align}

For the $\vphi^4$ model with $\lambda>0$, however, the last term on the
left-hand side of \refeq{schwinger-dyson} destroys linearity, and therefore it
is not obvious any more whether we can get an explicit expression for the
two-point function, or at least we can estimate its asymptotic behavior.

\subsection{The main result}
In this paper, we extend the lace-expansion methodology to factorize the
nonlinear term in \refeq{schwinger-dyson} and prove the expected asymptotic
behavior of the critical two-point function for $d>4$.  By virtue of this
approach, we can avoid the assumption of reflection positivity.  The precise
statement is the following.

\begin{thm}\label{thm:main}
For $d>4$, there is a $\lambda_0=\lambda_0(d,\sJ)\in(0,\infty)$ such
that the following holds for all $\lambda\in[0,\lambda_0]$: there is a
$\varPhi_\mu(x)=\Exp{\vphi_o^2}_\mu\delta_{o,x}+O(\lambda)/(|x|\vee1)^{3(d-2)}$
uniformly in $\mu>\muc$ such that a linearized version of the Schwinger-Dyson
equation
\begin{align}\lbeq{SDeq}
-\sum_v\sJ_{o,v}\Exp{\vphi_v\vphi_x}_\mu+\mu\Exp{\vphi_o\vphi_x}_\mu+\frac
 \lambda2\sum_v\varPhi_\mu(v)\Exp{\vphi_v\vphi_x}_\mu=\delta_{o,x}
\end{align}
holds.  Consequently,
\begin{align}\lbeq{mucthm}
\muc=\hat\sJ-\frac\lambda2\Exp{\vphi_o^2}_{\muc}+O(\lambda^2),
\end{align}
and there are $A=\hat\sJ V+O(\lambda^2)$ and $\kappa<2$ such that, as
$|x|\to\infty$,
\begin{align}\lbeq{2ptthm}
\Exp{\vphi_o\vphi_x}_{\muc}=\frac{\frac{d}2\Gamma(\frac{d-2}2)/\pi^{d/2}}
 {A|x|^{d-2}}+O(|x|^{\kappa-d}).
\end{align}
\end{thm}

\begin{rmk}
{\rm
\begin{enumerate}[(a)]
\item
We may prove similar results for arbitrarily large $\lambda$ if $\hat\sJ$ is
sufficiently large (e.g., the nearest-neighbor model with $d\gg4$).  In fact,
the $O(\lambda)$ term in the above $\varPhi_\mu$ is actually
$O(\lambda/\mu^3)$ (cf., \refeq{varPhiNdef}).  Although the constant in the
$O(\lambda/\mu^3)$ term may depend on the range of $\sJ$, which is potentially
large, the denominator $\mu^3$ ($\simeq\hat\sJ^3$ around the critical point)
should be large enough to cancel that effect.
\item
We may also extend the results to the case of power-law decaying spin-spin
coupling, $\sJ(x)\propto|x|^{-d-\alpha}$ for some $\alpha>0$.  However,
the variance $V$ in \refeq{Vdef} does not exist if $\alpha<2$.  In this case,
the underlying random walk is in the domain of attraction of $\alpha$-stable
motion, and the critical two-point function $\Exp{\vphi_o\vphi_x}_{\muc}$
should asymptotically be a multiple of $|x|^{\alpha-d}$ as $|x|\to\infty$,
in dimensions $d>2\alpha$.  See \cite{cs14} for more details.
\item
The actual proof of \refeq{mucthm}--\refeq{2ptthm} assuming \refeq{SDeq} goes
as follows.  First, we note that the sum of \refeq{SDeq} yields
\begin{align}\lbeq{chiid}
-\hat\sJ+\mu+\frac\lambda2\sum_v\varPhi_\mu(v)=\chi_\mu^{-1}.
\end{align}
Using this, we can rearrange \refeq{SDeq} as
\begin{align}
\mu\Exp{\vphi_o\vphi_x}_\mu&=\delta_{o,x}+\sum_v\frac{\sJ(v)-\frac\lambda2
 \varPhi_\mu(v)}\mu~\mu\Exp{\vphi_v\vphi_x}_\mu\nn\\
&=\delta_{o,x}+\sum_v\frac{\mu-\chi_\mu^{-1}}\mu\frac{\sJ(v)-\frac\lambda2
 \varPhi_\mu(v)}{\hat\sJ-\frac\lambda2\sum_y\varPhi_\mu(y)}~\mu\Exp{\vphi_v
 \vphi_x}_\mu.
\end{align}
Let
\begin{align}
\cG_\mu(x)=\mu\Exp{\vphi_o\vphi_x}_\mu,&&
\cD_\mu(x)=\frac{\sJ(x)-\frac\lambda2\varPhi_\mu(x)}{\hat\sJ-\frac\lambda2
 \sum_y\varPhi_\mu(y)},
\end{align}
so that
\begin{align}
\cG_\mu(x)=\delta_{o,x}+\sum_v\bigg(1-\frac{\chi_\mu^{-1}}\mu\bigg)\cD_\mu(v)\,
 \cG_\mu(x-v).
\end{align}
This is identical to the convolution equation for the random-walk's Green
function generated by the 1-step distribution $\cD_\mu$ with killing rate
$\chi_\mu^{-1}/\mu$.  Therefore, by the standard random-walk analysis
(e.g., \cite[Proposition~2.1]{cs14}), we obtain
\begin{align}\lbeq{cGasy}
\cG_{\muc}(x)=\frac{\frac{d}2\Gamma(\frac{d-2}2)/\pi^{d/2}}{\sum_y|y|^2\cD_{
 \muc}(y)}|x|^{2-d}+O(|x|^{\kappa-d}),
\end{align}
for some $\kappa<2$, where $\cD_{\muc}$ is defined in terms of arbitrary
subsequential limit
$\varPhi_{\muc}\equiv\lim_{\mu_j\downarrow\muc}\varPhi_{\mu_j}$, which exists
and obeys
\begin{align}
\varPhi_{\muc}(x)=\Exp{\vphi_o^2}_{\muc}\delta_{o,x}+\frac{O(\lambda)}
 {(|x|\vee1)^{3(d-2)}},
\end{align}
due to the uniformity of $\varPhi_\mu$ in $\mu>\muc$.  Using this and
\refeq{chiid}, we obtain
\begin{align}
\muc=\hat\sJ-\frac\lambda2\sum_v\varPhi_{\muc}(v)=\hat\sJ-\frac\lambda2
 \Exp{\vphi_o^2}_{\muc}+O(\lambda^2).
\end{align}
Moreover, since $3(d-2)=d+2+2(d-4)$, we obtain
\begin{align}
\sum_y|y|^2\cD_{\muc}(y)&=\frac1{\hat\sJ-\frac\lambda2\sum_y\varPhi_{\muc}(y)}
 \sum_{y\ne o}|y|^2\bigg(\sJ(y)-\frac\lambda2\varPhi_{\muc}(y)\bigg)\nn\\
&=\frac1{\muc}\bigg(\hat\sJ V+\sum_{y\ne o}\frac{O(\lambda^2)}{|y|^{d+2(d-4)}}
 \bigg)\equiv\frac{A}{\muc}.
\end{align}
This together with \refeq{cGasy} implies \refeq{2ptthm}.
\end{enumerate}
}
\end{rmk}

\subsection{Organization}
The rest of this paper is organized as follows.  In Section~\ref{s:Ising},
as a preliminary section, we introduce some notation and summarize relevant
properties of the two-point function.  Then, in Section~\ref{s:proof}, we
use those properties and the lace expansion for the Ising model to prove
the main theorem, Theorem~\ref{thm:main}.

\section{Approximation by the Ising model}\label{s:Ising}
In this section, we briefly review two key components for the proof of
Theorem~\ref{thm:main}.  One of them is the Griffiths-Simon construction
(Section~\ref{ss:GSconstr}), by which we can approximate the $\vphi^4$ model
with a sum of $N$ Ising systems.  The other component is the random-current
representation (Section~\ref{ss:RCrep}), by which we can think of the Ising
two-point function as a certain connectivity function.  As a result, we can
find many useful properties of the two-point function
(Section~\ref{ss:basics}).  The lace expansion for the Ising model
(Section~\ref{ss:laceexp}) is one of them.

\subsection{The Griffiths-Simon construction}\label{ss:GSconstr}
\begin{figure}[t]
\[ \includegraphics[scale=0.6]{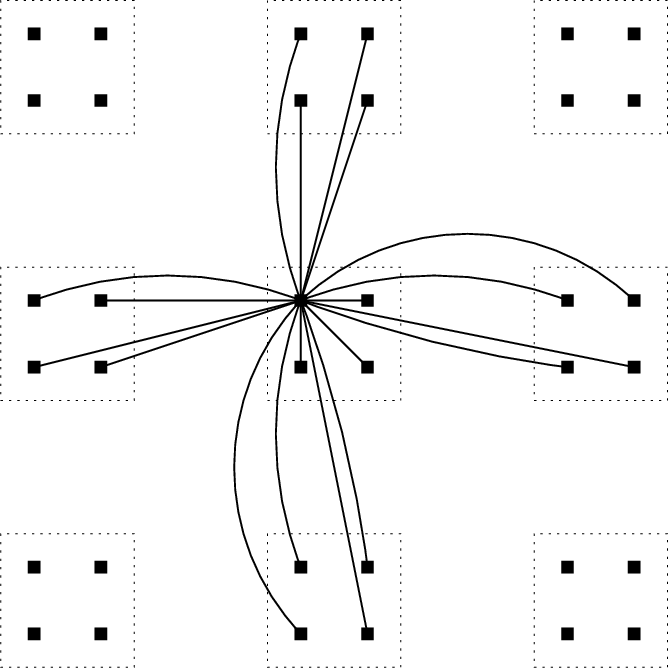} \]
\caption{\label{fig:Lambda[4]}The nearest-neighbor bonds from a single vertex
on $\tilde\mZ_4^d\equiv\Zd\times[4]$.  Each block contains four vertices
with a common spatial coordinate.}
\end{figure}
Let
\begin{align}
[N]=\{1,2,\dots,N\},&&
\tilde\Lambda_N=\Lambda\times[N],
\end{align}
and define the Ising Hamiltonian on $\tilde\Lambda_N$ as
\begin{align}\lbeq{IsingHamiltonian}
H_{\tilde\Lambda_N}(\sigmavec)&=-\frac12\sum_{\substack{x\ne y\in\Lambda\\
 i,j\in[N]}}J_{x,y}\sigma_{(x,i)}\sigma_{(y,j)}-\frac{I}2\sum_{
 \substack{x\in\Lambda\\ i,j\in[N]}}\sigma_{(x,i)}\sigma_{(x,j)}\nn\\
&=-\frac12\sum_{x\ne y\in\Lambda}J_{x,y}\wsigma_x\wsigma_y-\frac{I}2
 \sum_{x\in\Lambda}\wsigma_x^2,
\end{align}
where $\sigmavec\equiv(\sigma_{(x,i)})_{(x,i)\in\Lambda\times[N]}$ is the
Ising-spin configuration and
\begin{align}
\wsigma_x\equiv\sum_{i\in[N]}\sigma_{(x,i)}
\end{align}
is what we call in this paper the block spin at $x\in\Lambda$.
It is known \cite{sg73}  that, if $I,J$ and $\sigmavec$ are determined
from $\lambda,\mu,\sJ$ and $\vphivec$ with proper scaling (e.g.,
$\wsigma_x\mapsto\epsilon_N\wsigma_x$, with an appropriate scaling factor
$\epsilon_N$), then
\begin{align}\lbeq{GSlimit}
\epsilon_N^2\EXP{\wsigma_o\wsigma_x}_{\tilde\Lambda_N}\equiv\epsilon_N^2
 \frac{\dpst\sum_{\sigmavec\in\{\pm1\}^{\tilde\Lambda_N}}\wsigma_o\wsigma_x\,
 e^{-H_{\tilde\Lambda_N}(\sigmavec)}}{\dpst\sum_{\sigmavec\in\{\pm1\}^{\tilde
 \Lambda}}e^{-H_{\tilde\Lambda_N}(\sigmavec)}}\underset{N\uparrow\infty}\to
 \frac{\dpst\int_{\mR^\Lambda}\vphi_o\vphi_x\,e^{-\sH_\Lambda(\vphivec)}
 \,\text{d}\vphivec}{\dpst\int_{\mR^\Lambda}e^{-\sH_\Lambda(\vphivec)}\,
 \text{d}\vphivec}=\Exp{\vphi_o\vphi_x}_{\mu,\Lambda}.
\end{align}

Now, we provide heuristic explanation of the aforementioned proper scaling.
For more details, refer to \cite{sg73}.  First, we note that the marginal
distribution given $\wsigmavec=(\wsigma_x)_{x\in\Lambda}$ is
\begin{align}\lbeq{IsingPart1}
&\bigg(\frac12\bigg)^{|\tilde\Lambda_N|}\sum_{\substack{\sigmavec\in\{\pm1
 \}^{\tilde\Lambda_N}\\ (\wsigmavec\text{ fixed})}}e^{-H_{\tilde\Lambda_N}
 (\sigmavec)}\nn\\
&=\exp\bigg(\frac12\sum_{x\ne y\in\Lambda}J_{x,y}\wsigma_x\wsigma_y+\frac{I}2
 \sum_{x\in\Lambda}\wsigma_x^2\bigg)\prod_{x\in\Lambda}\sum_{\substack{
 \sigma_{(x,1)},\dots,\sigma_{(x,N)}\\ (\wsigma_x\text{ fixed})}}\bigg(\frac12
 \bigg)^N\nn\\
&=\exp\bigg(\frac12\sum_{x\ne y\in\Lambda}J_{x,y}\wsigma_x\wsigma_y+\frac{I}2
 \sum_{x\in\Lambda}\wsigma_x^2\bigg)\prod_{x\in\Lambda}\binom{N}{\frac{N+
 \wsigma_x}2}\bigg(\frac12\bigg)^N.
\end{align}
By Stirling's formula (i.e., $\sqrt{2\pi n}(\frac{n}e)^ne^{\frac1{12n+1}}\le
n!\le\sqrt{2\pi n}(\frac{n}e)^ne^{\frac1{12n}}$ for all $n\in\mN$),
\begin{align}
\log\Bigg(\binom{N}{\frac{N+\wsigma_x}2}\bigg(\frac12\bigg)^N\Bigg)
&=-\frac{N}2\Bigg(\bigg(1+\frac{\wsigma_x}N\bigg)\log\bigg(1+\frac{\wsigma_x}
 N\bigg)+\bigg(1-\frac{\wsigma_x}N\bigg)\log\bigg(1-\frac{\wsigma_x}N\bigg)
 \Bigg)\nn\\
&\quad+O(\wsigma_x^2/N^2)+\underbrace{O(\log N)}_{\text{independent of}
 ~\wsigma_x}.
\end{align}
Let $\eta_x=\wsigma_x/N$. Then, by the Taylor expansion,
\begin{align}
(1\pm\eta_x)\log(1\pm\eta_x)=(1\pm\eta_x)\bigg(\pm\eta_x-\frac{\eta_x^2}2
 \pm\frac{\eta_x^3}3-\frac{\eta_x^4}4\pm\frac{\eta_x^5}5+o(\eta_x^5)\bigg),
\end{align}
which implies
\begin{align}
(1+\eta_x)\log(1+\eta_x)+(1-\eta_x)\log(1-\eta_x)=2\bigg(\frac{\eta_x^2}2
 +\frac{\eta_x^4}{12}+o(\eta_x^5)\bigg).
\end{align}
Therefore,
\begin{align}\lbeq{IsingPart2}
\refeq{IsingPart1}\propto\exp\bigg(\frac12\sum_{x\ne y\in\Lambda}J_{x,y}
 \wsigma_x\wsigma_y\bigg)\prod_{x\in\Lambda}\exp\Bigg(\frac12\bigg(I-\frac1N
 +O(N^{-2})\bigg)\wsigma_x^2-\frac1{12}\frac{\wsigma_x^4}{N^3}+o\bigg(
 \frac{\wsigma_x^5}{N^4}\bigg)\Bigg).
\end{align}
Let
\begin{align}\lbeq{scaling1}
\frac1{12}\frac{\wsigma_x^4}{N^3}=\frac\lambda{4!}\vphi_x^4,\quad
 \text{or equivalently}\quad
 \vphi_x=\epsilon_N\wsigma_x\equiv
 \bigg(\frac\lambda2N^3\bigg)^{-1/4}\wsigma_x,
\end{align}
and
\begin{align}\lbeq{scaling2}
J_{x,y}=\sJ_{x,y}\epsilon_N^2,&&
I=\frac1N-\mu\epsilon_N^2.
\end{align}
Then, we arrive at
\begin{align}
\refeq{IsingPart2}&=\exp\Bigg(\frac12\sum_{x\ne y\in\Lambda}\sJ_{x,y}\vphi_x
 \vphi_y-\sum_{x\in\Lambda}\bigg(\frac{\mu+O(N^{-1/2})}2\vphi_x^2
 +\frac\lambda{4!}\vphi_x^4+o(N^{-1/4}\vphi_x^5)\bigg)\Bigg)\nn\\
&\sim e^{-\sH_\Lambda(\vphivec)}.
\end{align}

In Section~\ref{s:proof}, we apply the lace expansion \cite{s07} to
the ferromagnetic Ising model defined by the Hamiltonian
\refeq{IsingHamiltonian}.  For the ferromagnetic condition $I\ge0$,
we assume from now on
\begin{align}
N\ge\frac{2\mu^2}\lambda.
\end{align}

\subsection{The random-current representation}\label{ss:RCrep}
In this subsection, we explain the random-current representation (e.g.,
\cite{a82}) and introduce some notation.

First, we rewrite the Ising Hamiltonian \refeq{IsingHamiltonian} as
\begin{align}
H_{\tilde\Lambda_N}(\sigmavec)&=-\frac12\sum_{\substack{x\ne y\in\Lambda\\
 i,j\in[N]}}J_{x,y}\sigma_{(x,i)}\sigma_{(y,j)}-\frac{I}2\sum_{\substack{x
 \in\Lambda\\ i\ne j\in[N]}}\sigma_{(x,i)}\sigma_{(x,j)}-\frac{I}2|\tilde
 \Lambda_N|\nn\\
&=-\sum_{b\in\mB_{\tilde\Lambda_N}}\tilde J_b\sigma_{b_1}\sigma_{b_2}
 -\frac{I}2|\tilde\Lambda|,
\end{align}
where $\mB_{\tilde\Lambda_N}=\{b=\{b_1,b_2\}:b_1\ne b_2\in\tilde\Lambda_N\}$
and
\begin{align}
\tilde J_{(x,i),(y,j)}=J_{x,y}+I\delta_{x,y}(1-\delta_{i,j})=
 \begin{cases}
 J_{x,y}\quad&[x\ne y],\\
 I&[x=y\text{ and }i\ne j],\\
 0&[\text{otherwise}].
 \end{cases}
\end{align}
Then, by expanding exponentials, we obtain
\begin{align}
\bigg(\frac12\bigg)^{|\tilde\Lambda_N|}&\sum_{\sigmavec\in\{\pm1\}^{\tilde
 \Lambda}}e^{-H_{\tilde\Lambda_N}(\sigmavec)}=\bigg(\frac{e^{I/2}}2\bigg)^{
 |\tilde\Lambda_N|}\sum_{\sigmavec\in\{\pm1\}^{\tilde\Lambda_N}}\prod_{b\in
 \mB_{\tilde\Lambda_N}}\sum_{n_b\in\Zp}\frac{(\tilde J_b\sigma_{b_1}\sigma_{
 b_2})^{n_b}}{n_b!}\nn\\
&=e^{I|\tilde\Lambda_N|/2}\sum_{\nvec=(n_b)}
 \bigg(\underbrace{\prod_{b\in\mB_{\tilde\Lambda_N}}\frac{\tilde J_b^{n_b}}
 {n_b!}}_{w_{\tilde\Lambda_N}(\nvec)}\bigg)\prod_{\tilde x\in\tilde\Lambda_N}
 \underbrace{\frac12\sum_{\sigma_{\tilde x}\in\{\pm1\}}\sigma_{\tilde
 x}^{\sum_{b\ni\tilde x}n_b}}_{\ind{\sum_{b\ni\tilde x}n_b\text{ is even}}}
 \nn\\
&=e^{I|\tilde\Lambda_N|/2}\sum_{\partial\nvec=\vno}w_{\tilde\Lambda_N}(\nvec),
\end{align}
where $\partial\nvec\equiv\{\tilde x\in\tilde\Lambda_N:\sum_{b\ni\tilde x}n_b$
is odd$\}$ is the set of sources in the current configuration
$\nvec=(n_b)\in\Zp^{\mB_{\tilde\Lambda_N}}$.  Similarly, for
$\tilde x,\tilde y\in\tilde\Lambda_N$,
\begin{align}
\bigg(\frac12\bigg)^{|\tilde\Lambda_N|}\sum_{\sigmavec\in\{\pm1\}^{\tilde
 \Lambda}}\sigma_{\tilde x}\sigma_{\tilde y}\,e^{-H_{\tilde\Lambda_N}
 (\sigmavec)}=e^{I|\tilde\Lambda_N|/2}\sum_{\partial\nvec=\tilde x\vtri\tilde
 y}w_{\tilde\Lambda}(\nvec),
\end{align}
where $\tilde x\vtri\tilde y$ is the abbreviation for the symmetric difference
$\{\tilde x\}\triangle\{\tilde y\}$.  As a result, we arrive at the
random-current representation for the Ising two-point function
\begin{align}\lbeq{RCrep}
\EXP{\sigma_{\tilde x}\sigma_{\tilde y}}_{\tilde\Lambda_N}\equiv\frac{\dpst
 \sum_{\sigmavec\in\{\pm1\}^{\tilde\Lambda_N}}\sigma_{\tilde x}\sigma_{\tilde
 y}\,e^{-H_{\tilde\Lambda_N}(\sigmavec)}}{\dpst\sum_{\sigmavec\in\{\pm1\}^{
 \tilde\Lambda}}e^{-H_{\tilde\Lambda_N}(\sigmavec)}}=\frac{\dpst\sum_{\partial
 \nvec=\tilde x\vtri\tilde y}w_{\tilde\Lambda_N}(\nvec)}{\dpst\sum_{\partial
 \nvec=\vno}w_{\tilde\Lambda}(\nvec)}.
\end{align}

\subsection{Basic properties of the two-point function}\label{ss:basics}
In this subsection, we summarize the properties of the Ising two-point
function obtained from the random-current representation \refeq{RCrep}.

\begin{lmm}[(2.28) and (2.37) of \cite{cs14}]
Let $\Lambda\subset\Zd$ be the $d$-dimensional hypercube centered at the
origin $o\in\Zd$.  For any $I\ge0$, the following two inequalities hold:
\begin{enumerate}[(i)]
\item
For any $x\in\Lambda$,
\begin{align}\lbeq{lmm:a-priori}
\EXP{\sigma_{\tilde o}\sigma_{\tilde x}}_{\tilde\Lambda_N}-\delta_{\tilde o,
 \tilde x}\le\sum_{\tilde v\in\tilde\Lambda_N}(\tanh\tilde J_{\tilde o,\tilde
 v})\,\EXP{\sigma_{\tilde v}\sigma_{\tilde x}}_{\tilde\Lambda_N}.
\end{align}
\item
Suppose that the radius of $\Lambda$ is bigger than a given $\ell<\infty$.
Then, for $|x|>\ell$,
\begin{align}\lbeq{lmm:Simon-Lieb}
\EXP{\sigma_{\tilde o}\sigma_{\tilde x}}_{\tilde\Lambda_N}\le
 \sum_{\substack{\tilde u,\tilde v\in\tilde\Lambda_N\\ (|u|\le\ell<|v|)}}
 \EXP{\sigma_{\tilde o}\sigma_{\tilde u}}_{\tilde\Lambda_N}(\tanh J_{u,v})
 \EXP{\sigma_{\tilde v}\sigma_{\tilde x}}_{\tilde\Lambda_N}.
\end{align}
\end{enumerate}
\end{lmm}

\begin{prp}\label{prp:Gproperties}
Let
\begin{align}\lbeq{GLambda}
G_{\tilde\Lambda_N}(o,x)=\frac{1-(N-1)\tanh I}N\EXP{\wsigma_o\wsigma_x}_{\tilde
 \Lambda_N},
\end{align}
and denote its (unique and translation-invariant) infinite-volume limit by
\begin{align}\lbeq{GN}
G_N(x)=\lim_{\Lambda\uparrow\Zd}G_{\tilde\Lambda_N}(o,x).
\end{align}
Let $\mu>\mu_N\equiv\inf\{\mu:\sum_xG_N(x)<\infty\}.$
\begin{enumerate}[(i)]
\item
Let $S_p(x)$ be the random-walk's Green function whose 1-step distribution and
fugacity are defined, respectively, as
\begin{align}\lbeq{p&D}
D(v)=\frac{\tanh J_{o,v}}{\sum_{v\in\Zd}\tanh J_{o,v}},&&
p=\frac{N\sum_{v\in\Zd}\tanh J_{o,v}}{1-(N-1)\tanh I}.
\end{align}
Then, $G_N(x)\le S_p(x)$ for all $x\in\Zd$.
\item
Suppose that there is an $\alpha>0$ such that $J_{o,x}=O(|x|^{-d-\alpha})$ as
$|x|\to\infty$ ($\alpha$ can be an arbitrarily large number in the current
setting).  Then, there is a $K_\mu<\infty$ such that
$G_N(x)\le K_\mu(|x|\vee1)^{-d-\alpha}$ for all $x\in\Zd$.
\end{enumerate}
\end{prp}

\Proof{Proof of (i).}
Let $\tilde o=(o,i)$ and $\tilde x=(x,j)$.  Summing \refeq{lmm:a-priori} over
$i,j\in[N]$, we obtain
\begin{align}
\EXP{\wsigma_o\wsigma_x}_{\tilde\Lambda_N}-N\delta_{o,x}&\le\sum_{\tilde v\in
 \tilde\Lambda_N}\sum_{i\in[N]}(\tanh\tilde J_{(o,i),\tilde v})\,\EXP{\sigma_{
 \tilde v}\wsigma_x}_{\tilde\Lambda_N}\nn\\
&=\sum_{v\in\Lambda}\sum_{i,j\in[N]}\tanh\tilde J_{(o,i),(v,j)}\frac{\EXP{
 \wsigma_v\wsigma_x}_{\tilde\Lambda_N}}N\nn\\
&\le(N-1)(\tanh I)\EXP{\wsigma_o\wsigma_x}_{\tilde\Lambda_N}+N\sum_{v\in
 \Lambda}(\tanh J_{o,v})\EXP{\wsigma_v\wsigma_x}_{\tilde\Lambda_N}.
\end{align}
Solving this inequality for $\EXP{\wsigma_o\wsigma_x}_{\tilde\Lambda_N}$ and
using \refeq{GLambda}--\refeq{p&D}, we arrive at
\begin{align}
G_N(x)\le\delta_{o,x}+p(D*G_N)(x)\equiv
\delta_{o,x}+p\sum_{v\in\Zd}D(v)\,G_N(x-v).
\end{align}
Repeated application of this inequality yields
$G_N(x)\le\sum_{n=0}^\infty p^nD^{*n}(x)=S_p(x)$.
\QED

\Proof{Proof of (ii).}
Let $\tilde o=(o,i)$ and $\tilde x=(x,j)$.  Summing \refeq{lmm:Simon-Lieb}
over $i,j\in[N]$ and using \refeq{GLambda}--\refeq{p&D}, we readily obtain
the Simon-Lieb type inequality
\begin{align}
G_N(x)&\le\sum_{\substack{u,v\in\Zd\\ (|u|\le\ell<|v|)}}G_N(u)\,\frac{N\tanh
 J_{u,v}}{1-(N-1)\tanh I}\,G_N(x-v)\nn\\
&=\sum_{\substack{u,v\in\Zd\\ (|u|\le\ell<|v|)}}G_N(u)\,pD(v-u)\,G_N(x-v).
\end{align}
Under the assumption on the decay of $J$, the 1-step distribution $D$ obeys
the same asymptotic bound $D(x)=O(|x|^{-d-\alpha})$ as $|x|\to\infty$.
Then, we can follow the same proof as \cite[Lemma~2.4]{cs14} to obtain
$G_N(x)\le K_\mu(|x|\vee1)^{-d-\alpha}$, where $K_\mu$ is finite as long as
$\mu>\mu_N$.
\QED

Before closing this section, we provide bounds on
$\EXP{\sigma_{\tilde o}\sigma_{\tilde x}}_{\tilde\Lambda_N}$ in terms of
$G_N(x)$.  Notice that, by symmetry,
\begin{align}\lbeq{block-symmetry}
\EXP{\wsigma_o\wsigma_x}_{\tilde\Lambda_N}=N\EXP{\sigma_{\tilde o}
 \wsigma_x}_{\tilde\Lambda_N}=N\times
 \begin{cases}
 1+(N-1)\EXP{\sigma_{\tilde o}\sigma_{\tilde o'}}_{\tilde\Lambda_N}&[x=o],\\
 N\EXP{\sigma_{\tilde o}\sigma_{\tilde x}}_{\tilde\Lambda_N}&[x\ne o],
 \end{cases}
\end{align}
where $\tilde o'$ is another vertex than $\tilde o$ whose spatial coordinate
is $o$.  Then, for $\tilde x\ne\tilde o$,
\begin{align}\lbeq{block-symmetry-appl}
\EXP{\sigma_{\tilde o}\sigma_{\tilde x}}_{\tilde\Lambda_N}=\frac{\frac1N
 \EXP{\wsigma_o\wsigma_x}_{\tilde\Lambda_N}-\delta_{o,x}}{N-\delta_{o,x}}
&\le\frac{G_{\tilde\Lambda_N}(o,x)}{(1-(N-1)\tanh I)(N-1)}\nn\\
&\le\frac{G_{\tilde\Lambda_N}(o,x)}{\mu\epsilon_N^2(N-1)^2}.
\end{align}
Since $N^2/(N-1)^2\le4$ for $N\ge2$ and $G_{\tilde\Lambda_N}(o,x)\le G_N(x)$,
we have
\begin{align}\lbeq{bound-by-G1}
\EXP{\sigma_{\tilde o}\sigma_{\tilde x}}_{\tilde\Lambda_N}\le\delta_{\tilde o,
 \tilde x}+(1-\delta_{\tilde o,\tilde x})\frac{2G_N(x)}{\mu\epsilon_N^2N^2}.
\end{align}
Similarly, we obtain
\begin{align}\lbeq{bound-by-G2}
\sum_{\tilde v}(\tanh\tilde J_{\tilde o,\tilde v})\EXP{\sigma_{\tilde v}
 \sigma_{\tilde x}}_{\tilde\Lambda_N}&\le(\tanh I)\EXP{\wsigma_o\sigma_{\tilde
 x}}_{\tilde\Lambda_N}+\sum_v(\tanh J_{o,v})\EXP{\wsigma_v\sigma_{\tilde
 x}}_{\tilde\Lambda_N}\nn\\
&=\frac{(\tanh I)\,G_{\tilde\Lambda_N}(o,x)+\sum_v(\tanh J_{o,v})\,G_{\tilde
 \Lambda_N}(v,x)}{1-(N-1)\tanh I}\nn\\
&\le\frac1{\mu\epsilon_N^2N^2}\Big(G_N(x)+\hat\sJ\epsilon_N^2N(D*G_N)(x)
 \Big),
\end{align}
where the factor $\hat\sJ\epsilon_N^2N$ is smaller than 1 if we choose
$N>2\hat\sJ^2/\lambda$.

\section{Proof of the main theorem}\label{s:proof}
In this section, we prove Theorem~\ref{thm:main} by first showing the
expected $x$-space infrared bound (Section~\ref{ss:bounds}), which has been
proven to be true only for the nearest-neighbor model so far \cite{s82}.
Then, by using that infrared bound, we derive the linear Schwinger-Dyson
equation (Section~\ref{ss:LSD}), which is the core of the main theorem.
Both sections depend heavily on the lace expansion for the Ising model
(Section~\ref{ss:laceexp}).

\subsection{The lace expansion}\label{ss:laceexp}
The lace expansion has been successful in proving asymptotic behavior of
the critical two-point function for various models.  In particular,
for the ferromagnetic Ising model, which is considered to be in the same
universality class as the $\vphi^4$ model, the critical two-point function
is proven to be $|x|^{2-d}$ times a model-dependent constant as
$|x|\to\infty$ when the spin-spin coupling has a finite $(2+\vep)$th moment
for some $\vep>0$ \cite{cs14,s07}.

In this subsection, we apply the lace expansion for the Ising model
\cite{s07} to the approximate model constructed in Section~\ref{ss:GSconstr}
and investigate the $N$-dependence of the expansion coefficients.

According to \cite{s07}, for every $T\in\Zp$, there are functions
$\pi_{\tilde\Lambda_N}^{\sss(\le T)}$ and $r_{\tilde\Lambda_N}^{\sss(T+1)}$
on $\tilde\Lambda_N\times\tilde\Lambda_N$ such that the following identity
holds:
\begin{align}\lbeq{laceexp}
\EXP{\sigma_{\tilde o}\sigma_{\tilde x}}_{\tilde\Lambda_N}=\pi_{\tilde
 \Lambda_N}^{\sss(\le T)}(\tilde o,\tilde x)+\sum_{\tilde u,\tilde v\in\tilde
 \Lambda_N}\pi_{\tilde\Lambda_N}^{\sss(\le T)}(\tilde o,\tilde u)\,(\tanh
 \tilde J_{\tilde u,\tilde v})\,\EXP{\sigma_{\tilde v}\sigma_{\tilde
 x}}_{\tilde\Lambda_N}+r_{\tilde\Lambda_N}^{\sss(T+1)}(\tilde o,\tilde x).
\end{align}
In fact, $\pi_{\tilde\Lambda_N}^{\sss(\le T)}(\tilde o,\tilde x)$ is an
alternating sum of nonnegative functions
$\pi_{\tilde\Lambda_N}^{\sss(t)}(\tilde o,\tilde x)$, $0\le t\le T$.
Moreover, the remainder
$r_{\tilde\Lambda_N}^{\sss(T+1)}(\tilde o,\tilde x)$ is bounded uniformly in
$x$ as
\begin{align}\lbeq{rbd}
|r_{\tilde\Lambda_N}^{\sss(T+1)}(\tilde o,\tilde x)|&\le\sum_{\tilde u}
 \pi_{\tilde\Lambda_N}^{\sss(T)}(\tilde o,\tilde u)\sum_{\tilde v}(\tanh
 \tilde J_{\tilde u,\tilde v})\,\EXP{\sigma_{\tilde v}\sigma_{\tilde
 x}}_{\tilde\Lambda_N}\nn\\
&\le(1+\hat\sJ\epsilon_N^2N)\sum_{\tilde u}\pi_{\tilde
 \Lambda_N}^{\sss(T)}(\tilde o,\tilde u),
\end{align}
where we have used the inequality
\begin{align}
\sum_{\tilde v}\tanh\tilde J_{\tilde u,\tilde v}=(N-1)\tanh I+N\sum_v\tanh
 J_{u,v}\le1+\hat\sJ\epsilon_N^2N.
\end{align}

\begin{figure}[t]
\begin{gather*}
\pi_{\tilde\Lambda_N}^{\sss(0)}(\tilde o,\tilde x)\le{\sst\tilde o}\,
 \raisebox{-1.1pc}{\includegraphics[scale=0.3]{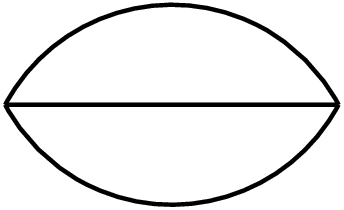}}\,{\sst\tilde x}\hskip4pc
\pi_{\tilde\Lambda_N}^{\sss(1)}(\tilde o,\tilde x)\le\sideset{_{\tilde
 o}}{_{\tilde x}}{\mathop{\raisebox{-1pc}{\includegraphics[scale=0.4]{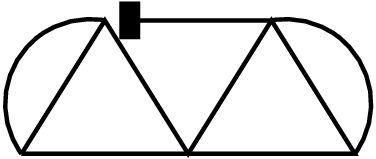}}}}
 +~\cdots\\[2pc]
\pi_{\tilde\Lambda_N}^{\sss(2)}(\tilde o,\tilde x)\le\sideset{_{\tilde o}}
 {^{\tilde x}}{\mathop{\raisebox{-1.1pc}{\includegraphics[scale=0.4]{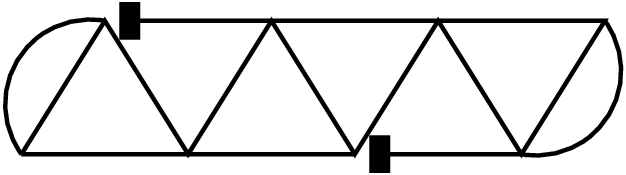}}}}
 +\sideset{_{\tilde o}}{^{\tilde x}}{\mathop{\raisebox{-1.1pc}
 {\includegraphics[scale=0.4]{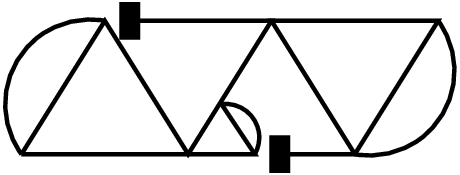}}}}+~\cdots
\end{gather*}
\caption{\label{fig:pi012}The leading bounding diagrams for
$\pi_{\tilde\Lambda_N}^{\sss(t)}(\tilde o,\tilde x)$, $t=0,1,2$.
Each line segment represents an Ising two-point function, e.g.,
$\pi_{\tilde\Lambda_N}^{\sss(0)}(\tilde o,\tilde x)\le
 \EXP{\sigma_{\tilde o}\sigma_{\tilde x}}_{\tilde\Lambda_N}^3$.
The unlabeled vertices are summer over $\tilde\Lambda_N$.
The tiny rectangles represent $\tanh\tilde J$.}
\end{figure}

The functions $\pi_{\tilde\Lambda_N}^{\sss(t)}(\tilde o,\tilde x)$, $t\ge0$,
are defined by using the random-current representation.  For example,
\begin{align}\lbeq{pi0def}
\pi_{\tilde\Lambda_N}^{\sss(0)}(\tilde o,\tilde x)=\frac{\dpst\sum_{\partial
 \nvec=\tilde o\vtri\tilde x}w_{\tilde\Lambda_N}(\nvec)\,\ind{\tilde o\underset
 {\nvec}{\db}\tilde x}}{\dpst\sum_{\partial\nvec=\vno}w_{\tilde\Lambda_N}
 (\nvec)},
\end{align}
where $\tilde o\underset{\nvec}{\db}\tilde x$ means that there are at least
two bond-disjoint paths in $\tilde\Lambda_N$ from $\tilde o$ to $\tilde x$,
consisting of bonds $b$ with $n_b>0$.  The precise definitions of those
functions are irrelevant, and we refrain from showing them here.  What
matters most is their diagrammatic bounds \cite[Proposition~4.1]{s07} (cf.,
Figure~\ref{fig:pi012}).
Combining with \refeq{bound-by-G1}--\refeq{bound-by-G2}, we can show the
following proposition.

\begin{prp}\label{prp:pibds}
Let
\begin{align}\lbeq{tlambdef}
\veee{x}=|x|\vee1,&&
\tlamb=\frac1{\mu\epsilon_N^2N^2}=\frac1\mu\sqrt{\frac\lambda{2N}},
\end{align}
and let $\lambda\ll2\mu^2$ and $N>2(\hat\sJ\vee\mu)^2/\lambda$ (so that
$\hat\sJ\epsilon_N^2N<1$ and $\tlamb<\tlamb^2N\ll1$).  Suppose that
$\sup_x\veee{x}^{d-2}G_N(x)$ is bounded by a constant which is
independent of $\lambda,\mu,N$.  Then, for $d>4$,
\begin{align}
&0\le\pi_{\tilde\Lambda_N}^{\sss(0)}(\tilde o,\tilde x)-\delta_{\tilde o,\tilde
 x}\le\frac{O(\tlamb)^3}{\veee{x}^{3(d-2)}},\lbeq{pi0bd}\\
&0\le\pi_{\tilde\Lambda_N}^{\sss(1)}(\tilde o,\tilde x)-\delta_{\tilde o,\tilde
 x}\sum_{\tilde v}(\tanh\tilde J_{\tilde o,\tilde v})\EXP{\sigma_{\tilde v}
 \sigma_{\tilde o}}_{\tilde\Lambda_N}\le O(\tlamb)^2\bigg(\delta_{\tilde o,
 \tilde x}+\frac{O(\tlamb)}{\veee{x}^{3(d-2)}}\bigg),\lbeq{pi1bd}\\
&0\le\pi_{\tilde\Lambda_N}^{\sss(t)}(\tilde o,\tilde x)\le O(\tlamb)^t\bigg(
 \delta_{\tilde o,\tilde x}+\frac{O(\tlamb)}{\veee{x}^{3(d-2)}}\bigg)\qquad
 [t\ge2],\lbeq{pige2bd}
\end{align}
where the constants in the $O(\tlamb)$ terms are independent of
$\lambda,\mu,N$ and $\Lambda$.
\end{prp}

As a result of the above proposition and \refeq{rbd}, we have
$\lim_{T\to\infty}r_{\tilde\Lambda_N}^{\sss(T+1)}(\tilde o,\tilde x)=0$, and
therefore the alternating series $\pi_{\tilde\Lambda_N}(\tilde o,\tilde x)
\equiv\lim_{T\to\infty}\pi_{\tilde\Lambda_N}^{\sss(\le T)}(\tilde o,\tilde x)$
converges and satisfies
\begin{align}\lbeq{pisumbd}
\bigg|\pi_{\tilde\Lambda_N}(\tilde o,\tilde x)-\delta_{\tilde o,\tilde x}
 \bigg(1-\sum_{\tilde v}(\tanh\tilde J_{\tilde o,\tilde v})\EXP{\sigma_{\tilde
 v}\sigma_{\tilde o}}_{\tilde\Lambda_N}\bigg)\bigg|\le O(\tlamb^2)\bigg(
 \delta_{\tilde o,\tilde x}+\frac{O(\tlamb)}{\veee{x}^{3(d-2)}}\bigg).
\end{align}
We will use this estimate in the next subsection to investigate (the
$T\to\infty$ limit of) \refeq{laceexp} and prove that the assumed bound
on $G_N$ in Proposition~\ref{prp:pibds} indeed holds.

\Proof{Proof of Proposition~\ref{prp:pibds}.}
The inequality \refeq{pi0bd} is readily obtained by applying \refeq{bound-by-G1}
to the diagrammatic bound $\pi_{\tilde\Lambda_N}^{\sss(0)}(\tilde o,\tilde x)\le
 \EXP{\sigma_{\tilde o}\sigma_{\tilde x}}_{\tilde\Lambda_N}^3$ in
\cite[Proposition~4.1]{s07} (see also \cite{s08} for intuitive explanation).
The proof of the other inequalities
\refeq{pi1bd}--\refeq{pige2bd} are much more involved, and we only explain
in detail how to bound the leading diagram for
$\pi_{\tilde\Lambda_N}^{\sss(1)}(\tilde o,\tilde x)$, which is depicted in
Figure~\ref{fig:pi012}:
\begin{align}\lbeq{pi1leading}
\pi_{\tilde\Lambda_N}^{\sss(1)}(\tilde o,\tilde x)\le\sum_{\tilde u,\tilde v,
 \tilde y}{\sum_{\tilde w}}'&\EXP{\sigma_{\tilde o}\sigma_{\tilde u}}_{
 \tilde\Lambda_N}^2\EXP{\sigma_{\tilde o}\sigma_{\tilde w}}_{\tilde\Lambda_N}
 \EXP{\sigma_{\tilde w}\sigma_{\tilde u}}_{\tilde\Lambda_N}(\tanh\tilde J_{
 \tilde u,\tilde v})\EXP{\sigma_{\tilde v}\sigma_{\tilde y}}_{\tilde\Lambda_N}
 \nn\\
&\times\EXP{\sigma_{\tilde y}\sigma_{\tilde w}}_{\tilde\Lambda_N}
 \EXP{\sigma_{\tilde w}\sigma_{\tilde x}}_{\tilde\Lambda_N}
 \EXP{\sigma_{\tilde y}\sigma_{\tilde x}}_{\tilde\Lambda_N}^2
 +\text{ error term},
\end{align}
where $\sum'_{\tilde w}$ is interpreted as the sum over the singleton
$\{\tilde o\}$ if $\tilde u=\tilde o$, or over the singleton $\{\tilde x\}$ if
$\tilde y=\tilde x$, or over $\tilde\Lambda_N$ otherwise.  The leading term is
the contribution from $P_{\Lambda;u}^{\prime\sss(0)}$ in \cite[(4.12)]{s07},
while the error term is the contribution from the series
$\sum_{j\ge1}P_{\Lambda;u}^{\prime\sss(j)}$ in \cite[(4.12)]{s07}.
By a simpler version of the diagrammatic bounds explained below, we can
show that $P_{\Lambda;u}^{\prime\sss(j)}$ for $j\ge1$ is bounded as
\cite[(5.14)]{s07}, which is the bound on $P_{\Lambda;u}^{\prime\sss(0)}$
in \cite[(5.12)]{s07} multiplied by the exponentially small factor
$O(\theta_0^2)^j$ (in the current setting, $\theta_0=\tlamb$).  Therefore,
we only need to bound the leading term in \refeq{pi1leading}.  The higher
order functions $\pi_{\tilde\Lambda_N}^{\sss(t)}(\tilde o,\tilde x)$,
$t\ge2$, can be estimated similarly; each extra factor
$\sum_{b:\underline{b}=y}\tau_bQ''_{\Lambda;u,v}(\bar b,x)$ in
\cite[(4.15)]{s07} gives rise to the $O(\theta_0)$ term in \cite[(5.17)]{s07},
which leads to the exponentially decaying bound in \refeq{pige2bd}.  For those
interested in more details about diagrammatic bounds, refer also to
\cite[Appendix~B]{hhs08} and \cite[Section~4]{s07}.

Now, we prove that the sum on the right-hand side of \refeq{pi1leading} obeys
the inequality \refeq{pi1bd}.  First, we split it into three sums depending on
whether (i)~$\tilde o=\tilde u$ and $\tilde y=\tilde x$ (hence
$\tilde w=\tilde o=\tilde x$), (ii)~$\tilde o\ne\tilde u$ and
$\tilde y=\tilde x$ (hence $\tilde w=\tilde x$), or
(iii)~$\tilde y\ne\tilde x$.  Then, we obtain
\begin{align}\lbeq{pi1decomp}
\text{The sum in \refeq{pi1leading}}&\le\delta_{\tilde o,\tilde x}\sum_{\tilde
 v}(\tanh\tilde J_{\tilde o,\tilde v})\EXP{\sigma_{\tilde v}\sigma_{\tilde
 o}}_{\tilde\Lambda_N}\nn\\
&\quad+\EXP{\sigma_{\tilde o}\sigma_{\tilde x}}_{\tilde\Lambda_N}\sum_{\tilde
 u(\ne\tilde o),\tilde v}\EXP{\sigma_{\tilde o}\sigma_{\tilde u}}_{\tilde
 \Lambda_N}^2\EXP{\sigma_{\tilde x}\sigma_{\tilde u}}_{\tilde\Lambda_N}(\tanh
 \tilde J_{\tilde u,\tilde v})\EXP{\sigma_{\tilde v}\sigma_{\tilde x}}_{\tilde
 \Lambda_N}\nn\\
&\quad+\sum_{\tilde u,\tilde v,\tilde w,\tilde y(\ne\tilde x)}\EXP{\sigma_{
 \tilde o}\sigma_{\tilde u}}_{\tilde\Lambda_N}^2\EXP{\sigma_{\tilde o}\sigma_{
 \tilde w}}_{\tilde\Lambda_N}\EXP{\sigma_{\tilde w}\sigma_{\tilde u}}_{\tilde
 \Lambda_N}(\tanh\tilde J_{\tilde u,\tilde v})\EXP{\sigma_{\tilde v}\sigma_{
 \tilde y}}_{\tilde\Lambda_N}\nn\\
&\hskip6pc\times\EXP{\sigma_{\tilde y}\sigma_{\tilde w}}_{\tilde\Lambda_N}
 \EXP{\sigma_{\tilde w}\sigma_{\tilde x}}_{\tilde\Lambda_N}
 \EXP{\sigma_{\tilde y}\sigma_{\tilde x}}_{\tilde\Lambda_N}^2.
\end{align}
In fact, the first term on the right-hand side is the trivial contribution to
$\pi_{\tilde\Lambda_N}^{\sss(1)}(\tilde o,\tilde x)$, and therefore
$\pi_{\tilde\Lambda_N}^{\sss(1)}(\tilde o,\tilde x)-\delta_{\tilde o,\tilde x}
\sum_{\tilde v}(\tanh\tilde J_{\tilde o,\tilde v})\EXP{\sigma_{\tilde v}
\sigma_{\tilde o}}_{\tilde\Lambda_N}\ge0$.

It remains to show that the second and third terms on the right-hand side of
the above inequality are bounded by the right-hand side of \refeq{pi1bd}.
In order to achieve this goal, we use the following convolution bounds.

\begin{lmm}[\cite{cs14,hhs03}]\label{lmm:convbds}
\begin{enumerate}[(i)]
\item
For any $a\ge b>0$ with $a\ne d$ and $a+b>d$, there is a $C<\infty$ such that
\begin{align}\lbeq{conv1}
\sum_{y\in\Zd}\veee{x-y}^{-a}\,\veee{y}^{-b}\le C\veee{x}^{a\vee d-a-b}.
\end{align}
\item
Let $f$ and $g$ be functions on $\Zd$, with $g$ being $\Zd$-symmetric.
Suppose that there are $C_1,C_2,C_3>0$ and $\rho>0$ such that
\begin{align*}
f(x)=C_1\veee{x}^{2-d},&&
|g(x)|\le C_2\delta_{o,x}+C_3\veee{x}^{-d-\rho}.
\end{align*}
Then there is a $\rho'\in(0,\rho\wedge2)$ such that, for $d>2$,
\begin{align}\lbeq{conv2}
(f*g)(x)=\frac{C_1\|g\|_1}{\veee{x}^{d-2}}+\frac{O(C_1C_3)}
 {\veee{x}^{d-2+\rho'}}.
\end{align}
\end{enumerate}
\end{lmm}

\begin{figure}
\begin{align*}
&\left.\begin{array}{l}
  \refeq{0delta-convbd}:~\sideset{_{\tilde x_1}^{\tilde x_3}}{_{\tilde x_2}^{
  \tilde x_4}}{\mathop{~\raisebox{-0.8pc}{\includegraphics[scale=0.3]{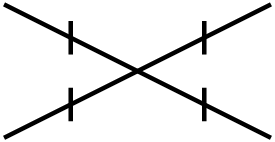}}~}}
  \\[2pc]
  \refeq{1delta-convbd}:~\sideset{_{\tilde x_1}^{\tilde x_3}}{_{\tilde x_2}^{
  \tilde x_4}}{\mathop{~\raisebox{-0.8pc}{\includegraphics[scale=0.3]{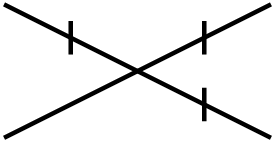}}~}}
 \end{array}\right\}
 \le~O(\tlamb^2N)~\sideset{_{\tilde x_1}^{\tilde x_3}}{_{\tilde x_2}^{\tilde
 x_4}}{\mathop{~\raisebox{-0.6pc}{\includegraphics[scale=0.3]{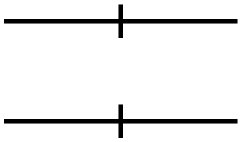}}~}}\\[1pc]
&~~\refeq{2delta-convbd1}:~\sideset{_{\tilde x_1}^{\tilde x_3}}{_{\tilde x_2}^{
 \tilde x_4}}{\mathop{~\raisebox{-0.8pc}{\includegraphics[scale=0.3]{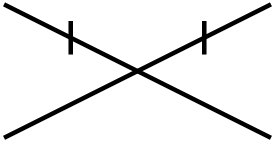}}~}}
 ~\le~O(\tlamb^2N)~\sideset{_{\tilde x_1}^{\tilde x_3}}{_{\tilde x_2}^{\tilde
 x_4}}{\mathop{~\raisebox{-0.5pc}{\includegraphics[scale=0.3]{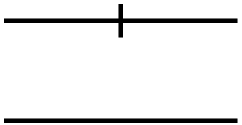}}~}}
 \\[1pc]
&~~\refeq{2delta-convbd2}:~\sideset{_{\tilde x_1}^{\tilde x_3}}{_{\tilde x_2}^{
 \tilde x_4}}{\mathop{~\raisebox{-0.8pc}{\includegraphics[scale=0.3]{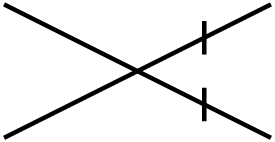}}~}}
 ~\le~O(1)~\sideset{_{\tilde x_1}^{\tilde x_3}}{_{\tilde x_2}^{\tilde x_4}}
 {\mathop{~\raisebox{-0.6pc}{\includegraphics[scale=0.3]{2l0d}}~}}\\[1pc]
&~~\refeq{3delta-convbd}:~\sideset{_{\tilde x_1}^{\tilde x_3}}{_{\tilde x_2}^{
 \tilde x_4}}{\mathop{~\raisebox{-0.8pc}{\includegraphics[scale=0.3]{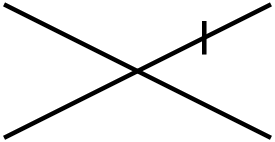}}~}}
 ~\le~O(1)~\sideset{_{\tilde x_1}^{\tilde x_3}}{_{\tilde x_2}^{\tilde x_4}}
 {\mathop{~\raisebox{-0.5pc}{\includegraphics[scale=0.3]{2l1d}}~}}\\[1pc]
&~~\refeq{4delta-convbd}:~\sideset{_{\tilde x_1}^{\tilde x_3}}{_{\tilde x_2}^{
 \tilde x_4}}{\mathop{~\raisebox{-0.8pc}{\includegraphics[scale=0.3]{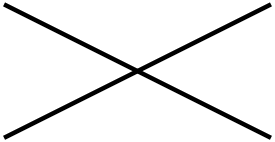}}~}}
 ~\le~O(1)~\sideset{_{\tilde x_1}^{\tilde x_3}}{_{\tilde x_2}^{\tilde x_4}}
 {\mathop{~\raisebox{-0.4pc}{\includegraphics[scale=0.3]{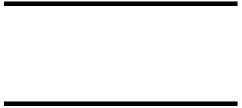}}~}}
\end{align*}
\caption{\label{fig:convbds}Schematic representations for
\refeq{0delta-convbd}--\refeq{4delta-convbd}.  If a line segment between
$\tilde u=(u,\cdot)$ and $\tilde v=(v,\cdot)$ is slashed, then it represents
$\tlamb/\veee{u-v}^{d-2}$; if it is unslashed, then it represents
$\delta_{\tilde u,\tilde v}+\tlamb/\veee{u-v}^{d-2}$.}
\end{figure}

We use Lemma~\ref{lmm:convbds}(i) to control the sums over
$\tilde u,\tilde w,\tilde y\in\Lambda_N$ in \refeq{pi1leading},
which correspond to the unlabeled vertices of degree 4 in the bounding diagram
in Figure~\ref{fig:pi012}.  For example, by Lemma~\ref{lmm:convbds}(i) with
$a=b=d-2$, we obtain that, for $d>4$,
\begin{align}\lbeq{0delta-convbd}
&\sum_{\tilde v\in\tilde\Lambda_N}\frac{\tlamb}{\veee{x_1-v}^{d-2}}\frac{\tlamb}
 {\veee{v-x_2}^{d-2}}\frac{\tlamb}{\veee{x_3-v}^{d-2}}\frac{\tlamb}{\veee{v-
 x_4}^{d-2}}\nn\\
&=N\sum_{v\in\Zd}\frac{\tlamb}{(\veee{x_1-v}\vee\veee{v-x_2})^{d-2}}\frac{
 \tlamb}{(\veee{x_1-v}\wedge\veee{v-x_2})^{d-2}}\nn\\
&\qquad\times\frac{\tlamb}{(\veee{x_3-v}\vee\veee{v-x_4})^{d-2}}\frac{\tlamb}
 {(\veee{x_3-v}\wedge\veee{v-x_4})^{d-2}}\nn\\
&\le4^{d-2}C\tlamb^2N\frac{\tlamb}{\veee{x_1-x_2}^{d-2}}\frac{\tlamb}{\veee{
 x_3-x_4}^{d-2}},
\end{align}
where we have used the triangle inequality
$\veee{x_i-v}\vee\veee{v-x_j}\ge\veee{x_i-x_j}/2$ (cf.,
Figure~\ref{fig:convbds}).
Similarly, if Kronecker's delta is added to one of those four fractions,
then we have
\begin{align}\lbeq{1delta-convbd}
&\sum_{\tilde v\in\tilde\Lambda_N}\bigg(\delta_{\tilde x_1,\tilde v}+\frac{
 \tlamb}{\veee{x_1-v}^{d-2}}\bigg)\frac{\tlamb}{\veee{v-x_2}^{d-2}}\frac{\tlamb}
 {\veee{x_3-v}^{d-2}}\frac{\tlamb}{\veee{v-x_4}^{d-2}}\nn\\
&\le\frac{\tlamb}{\veee{x_1-x_2}^{d-2}}\frac{2^{d-2}\tlamb^2}{\veee{x_3-x_4}^{d
 -2}}+4^{d-2}C\tlamb^2N\frac{\tlamb}{\veee{x_1-x_2}^{d-2}}\frac{\tlamb}{\veee{
 x_3-x_4}^{d-2}}\nn\\
&\le C'\tlamb^2N\frac{\tlamb}{\veee{x_1-x_2}^{d-2}}\frac{\tlamb}{\veee{x_3
 -x_4}^{d-2}},
\end{align}
with $C'=2^{d-2}+4^{d-2}C$, where we have used the assumption
$\tlamb<\tlamb^2N$.  Moreover, if there are two fractions with Kronecker's
delta, then we have
\begin{align}\lbeq{2delta-convbd1}
&\sum_{\tilde v\in\tilde\Lambda_N}\bigg(\delta_{\tilde x_1,\tilde v}+\frac{
 \tlamb}{\veee{x_1-v}^{d-2}}\bigg)\bigg(\delta_{\tilde v,\tilde x_2}+\frac{
 \tlamb}{\veee{v-x_2}^{d-2}}\bigg)\frac{\tlamb}{\veee{x_3-v}^{d-2}}\frac{\tlamb}
 {\veee{v-x_4}^{d-2}}\nn\\
&\le\bigg(\delta_{\tilde x_1,\tilde x_2}+\frac{\tlamb}{\veee{x_1-x_2}^{d-2}}
 \bigg)\frac{2^{d-2}\tlamb^2}{\veee{x_3-x_4}^{d-2}}+C'\tlamb^2N\frac{\tlamb}
 {\veee{x_1-x_2}^{d-2}}\frac{\tlamb}{\veee{x_3-x_4}^{d-2}}\nn\\
&\le(2^{d-2}+C')\tlamb^2N\bigg(\delta_{\tilde x_1,\tilde x_2}+\frac{\tlamb}
 {\veee{x_1-x_2}^{d-2}}\bigg)\frac{\tlamb}{\veee{x_3-x_4}^{d-2}},
\end{align}
or
\begin{align}\lbeq{2delta-convbd2}
&\sum_{\tilde v\in\tilde\Lambda_N}\bigg(\delta_{\tilde x_1,\tilde v}+\frac
 {\tlamb}{\veee{x_1-v}^{d-2}}\bigg)\frac{\tlamb}{\veee{v-x_2}^{d-2}}\bigg(
 \delta_{\tilde x_3,\tilde v}+\frac{\tlamb}{\veee{x_3-v}^{d-2}}\bigg)\frac{
 \tlamb}{\veee{v-x_4}^{d-2}}\nn\\
&\le\bigg(\delta_{\tilde x_1,\tilde x_3}+\frac{\tlamb}{\veee{x_1-x_3}^{d-2}}
 \bigg)\frac{\tlamb}{\veee{x_3-x_2}^{d-2}}\frac{\tlamb}{\veee{x_3-x_4}^{d-2}}
 \nn\\
&\quad+C'\tlamb^2N\frac{\tlamb}{\veee{x_1-x_2}^{d-2}}\frac{\tlamb}{\veee{x_3
 -x_4}^{d-2}}\nn\\
&\le\Big(1+(2^{d-2}+C')\tlamb^2N\Big)\frac{\tlamb}{\veee{x_1-x_2}^{d-2}}\frac{
 \tlamb}{\veee{x_3-x_4}^{d-2}}.
\end{align}
By similar computations, we can show that there is a $C''<\infty$ such that
\begin{align}\lbeq{3delta-convbd}
\sum_{\tilde v\in\tilde\Lambda_N}\prod_{j=1}^3\bigg(\delta_{\tilde x_j,\tilde
 v}+\frac{\tlamb}{\veee{x_j-v}^{d-2}}\bigg)\frac{\tlamb}{\veee
 {v-x_4}^{d-2}}&\le C''\bigg(\delta_{\tilde x_1,\tilde x_2}+\frac
 {\tlamb}{\veee{x_1-x_2}^{d-2}}\bigg)\frac{\tlamb}{\veee{x_3-x_4}^{d-2}},
\end{align}
and
\begin{align}\lbeq{4delta-convbd}
\sum_{\tilde v\in\tilde\Lambda_N}\prod_{j=1}^4\bigg(\delta_{\tilde x_j,\tilde
 v}+\frac{\tlamb}{\veee{x_j-v}^{d-2}}\bigg)&\le C''\bigg(\delta_{
 \tilde x_1,\tilde x_2}+\frac{\tlamb}{\veee{x_1-x_2}^{d-2}}\bigg)\bigg(\delta_{\tilde x_3,\tilde x_4}+\frac{\tlamb}{\veee{x_3-
 x_4}^{d-2}}\bigg).
\end{align}

\begin{figure}
\[ \sideset{_{\tilde o}}{_{\tilde x}}{\mathop{\raisebox{-1pc}{\includegraphics
 [scale=0.4]{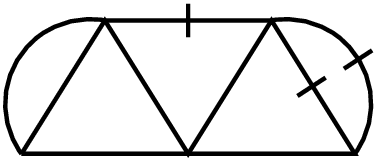}}}}^{\hskip3pc\sss(\tilde y)}~\stackrel
 {\refeq{1delta-convbd}}\lesssim~\sideset{_{\tilde o}}{^{\tilde x}}{\mathop
 {\raisebox{-1pc}{\includegraphics[scale=0.4]{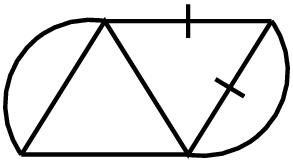}}}}_{\quad~\sss
 (\tilde w)}~\stackrel{\refeq{3delta-convbd}}\lesssim~\sideset{_{\tilde o}}
 {_{\tilde x}}{\mathop{\raisebox{-1pc}{\includegraphics[scale=0.4]
 {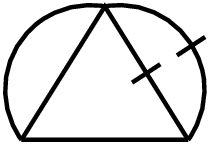}}}}^{\sss(\tilde u)}~\stackrel{\refeq{2delta-convbd2}}\lesssim
 ~{\sst\tilde o}\,\raisebox{-1pc}{\includegraphics[scale=0.4]{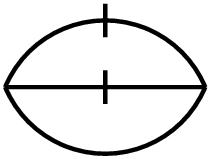}}\,
 {\sst\tilde x} \]
\caption{\label{fig:pi1decomp3bd}Schematic representations for
\refeq{pi1decomp3bd1}--\refeq{pi1decomp3bd2}.  Each inequality gives a
bound on the sum over vertices labeled with the letter in the parentheses.
The factors $O(\tlamb^2N)$ are omitted for brevity.}
\end{figure}

Now, we resume the proof of bounding the second and third terms on the
right-hand side of \refeq{pi1decomp}.  First, by \refeq{bound-by-G2} and
$\hat\sJ\epsilon_N^2N<1$, we have
\begin{align}
\sum_{\tilde v}(\tanh\tilde J_{\tilde u,\tilde v})\EXP{\sigma_{\tilde v}
 \sigma_{\tilde x}}_{\tilde\Lambda_N}\le\frac{O(\tlamb)}{\veee{x-u}^{d-2}}.
\end{align}
Therefore, by \refeq{1delta-convbd}, the second term on the right-hand side
of \refeq{pi1decomp} is bounded as
\begin{align}\lbeq{pi1decomp2bd}
&\EXP{\sigma_{\tilde o}\sigma_{\tilde x}}_{\tilde\Lambda_N}\sum_{\tilde u(\ne
 \tilde o),\tilde v}\EXP{\sigma_{\tilde o}\sigma_{\tilde u}}_{\tilde\Lambda_N}^2
 \EXP{\sigma_{\tilde x}\sigma_{\tilde u}}_{\tilde\Lambda_N}(\tanh\tilde
 J_{\tilde u,\tilde v})\EXP{\sigma_{\tilde v}\sigma_{\tilde x}}_{\tilde
 \Lambda_N}\nn\\
&\le\EXP{\sigma_{\tilde o}\sigma_{\tilde x}}_{\tilde\Lambda_N}\sum_{\tilde u}
 \frac{O(\tlamb)^2}{\veee{u}^{2(d-2)}}\bigg(\delta_{\tilde x,\tilde
 u}+\frac{O(\tlamb)}{\veee{x-u}^{d-2}}\bigg)\frac{O(\tlamb)}
 {\veee{x-u}^{d-2}}\nn\\
&\le\bigg(\delta_{\tilde o,\tilde x}+\frac{O(\tlamb)}{\veee{x}^{d-2}}\bigg)
 \frac{O(\tlamb^4N)}{\veee{x}^{2(d-2)}}.
\end{align}
Similarly, by \refeq{1delta-convbd}, the third term on the right-hand side of
\refeq{pi1decomp} is bounded as (cf., Figure~\ref{fig:pi1decomp3bd})
\begin{align}\lbeq{pi1decomp3bd1}
&\sum_{\tilde u,\tilde w}\EXP{\sigma_{\tilde o}\sigma_{\tilde u}}_{\tilde
 \Lambda_N}^2\EXP{\sigma_{\tilde o}\sigma_{\tilde w}}_{\tilde\Lambda_N}
 \EXP{\sigma_{\tilde w}\sigma_{\tilde u}}_{\tilde\Lambda_N}\EXP{\sigma_{\tilde
 w}\sigma_{\tilde x}}_{\tilde\Lambda_N}\nn\\
&\qquad\times\sum_{\tilde v,\tilde y(\ne\tilde x)}(\tanh\tilde J_{\tilde u,
 \tilde v})\EXP{\sigma_{\tilde v}\sigma_{\tilde y}}_{\tilde\Lambda_N}
 \EXP{\sigma_{\tilde y}\sigma_{\tilde w}}_{\tilde\Lambda_N}
 \EXP{\sigma_{\tilde y}\sigma_{\tilde x}}_{\tilde\Lambda_N}^2\nn\\
&\le\sum_{\tilde u,\tilde w}\EXP{\sigma_{\tilde o}\sigma_{\tilde u}}_{\tilde
 \Lambda_N}^2\EXP{\sigma_{\tilde o}\sigma_{\tilde w}}_{\tilde\Lambda_N}
 \EXP{\sigma_{\tilde w}\sigma_{\tilde u}}_{\tilde\Lambda_N}\EXP{\sigma_{\tilde
 w}\sigma_{\tilde x}}_{\tilde\Lambda_N}\nn\\
&\qquad\times\sum_{\tilde y}\frac{O(\tlamb)}{\veee{y-u}^{d-2}}
 \bigg(\delta_{\tilde w,\tilde y}+\frac{O(\tlamb)}{\veee{y-w}^{d-2}}\bigg)
 \frac{O(\tlamb)^2}{\veee{x-y}^{2(d-2)}}\nn\\
&\le\sum_{\tilde u}\EXP{\sigma_{\tilde o}\sigma_{\tilde u}}_{\tilde\Lambda_N}^2
 \frac{O(\tlamb^3N)}{\veee{x-u}^{d-2}}\sum_{\tilde w}\EXP{\sigma_{\tilde o}
 \sigma_{\tilde w}}_{\tilde\Lambda_N}\EXP{\sigma_{\tilde w}\sigma_{\tilde u}
 }_{\tilde\Lambda_N}\EXP{\sigma_{\tilde w}\sigma_{\tilde x}}_{\tilde\Lambda_N}
 \frac{O(\tlamb)}{\veee{x-w}^{d-2}}.
\end{align}
Then, by applying \refeq{3delta-convbd} to control the sum over $\tilde w$,
and then applying \refeq{2delta-convbd2} to control the sum over $\tilde u$,
we obtain
\begin{align}\lbeq{pi1decomp3bd2}
\refeq{pi1decomp3bd1}&\le\bigg(\delta_{\tilde o,\tilde x}+\frac{O(\tlamb)}
 {\veee{x}^{d-2}}\bigg)\sum_{\tilde u}\EXP{\sigma_{\tilde o}
 \sigma_{\tilde u}}_{\tilde\Lambda_N}^2\frac{O(\tlamb^4N)}{\veee{x-u}^{2(d
 -2)}}\nn\\
&\le\bigg(\delta_{\tilde o,\tilde x}+\frac{O(\tlamb)}{\veee{x}^{d-2}}\bigg)
 \frac{O(\tlamb^4N)}{\veee{x}^{2(d-2)}}.
\end{align}

Summarizing \refeq{pi1leading}--\refeq{pi1decomp} and
\refeq{pi1decomp2bd}--\refeq{pi1decomp3bd2} and then using $\tlamb^2N\ll1$, we
conclude that the sum on the right-hand side of \refeq{pi1leading} obeys the
inequality \refeq{pi1bd}.  We finish the proof of Proposition~\ref{prp:pibds}.
\QED

\subsection{Bound on the two-point function}\label{ss:bounds}
In this subsection, we prove that the assumed bound on $G_N$ in
Proposition~\ref{prp:pibds} indeed holds for all $\mu\in(\mu_N,\muNG]$
(n.b., $\mu_N\le\muNG$ is due to Proposition~\ref{prp:Gproperties}(i)), where
\begin{align}
\mu_N&=\inf\bigg\{\mu:\sum_{x\in\Zd}G_N(x)<\infty\bigg\},\\
\muNG&=\frac1{\epsilon_N^2}\bigg(\frac1N-\tanh^{-1}\frac{1-N\sum_x\tanh
 J_{o,x}}{N-1}\bigg).
\end{align}
We note that $\muNG$ is the value of $\mu$ at which $p=1$.  Although the exact
expression for $\muNG$ is unimportant, it shows that $\muNG$ tends to the
massless point $\hat\sJ$ for the Gaussian model as $N\to\infty$.  In fact,
\begin{align}\lbeq{muNGbd}
\muNG<\frac1{\epsilon_N^2}\bigg(\frac1N-\frac{1-N\sum_x\tanh J_{o,x}}{N-1}
 \bigg)<\frac{\hat\sJ N}{N-1}.
\end{align}
For now, we assume $\hat\sJ/2<\mu_N<\muNG$, which is to be verified later.

Let
\begin{align}\lbeq{tKdef}
\tK=\sup_{x\in\Zd}\veee{x}^{d-2}S_1(x),
\end{align}
where we recall that $S_1$ is the random-walk's Green function generated by
the 1-step distribution $D(x)=\tanh J_{o,x}/\sum_y\tanh J_{o,y}$.  Notice
that, under the assumption in Proposition~\ref{prp:pibds}
(cf., $\hat\sJ\epsilon_N^2N<1$ and $N\ge2$),
\begin{align}
J_{o,x}\ge\tanh J_{o,x}\ge J_{o,x}\bigg(1-\frac{J_{o,x}^2}3\bigg)\ge J_{o,x}
 \bigg(1-\frac{(\hat\sJ\epsilon_N^2)^2}3\bigg)\ge\frac{11}{12}J_{o,x},
\end{align}
and therefore $D$ inherits all the properties of $\sJ/\hat\sJ$.
In particular, there is a $\tau\in(0,2)$ such that
\begin{gather}
1-\hat D(k)=\frac{|k|^2}{2d}\sum_x|x|^2D(x)+O(|k|^{2+\tau}),\lbeq{1-hatDasy}\\
D^{*n}(x)\le\frac{O(n)}{\veee{x}^{d+2}},\lbeq{D*n1}\\
\bigg|D^{*n}(x)-\frac{D^{*n}(x+y)+D^{*n}(x-y)}2\bigg|\le\frac{O(n)|y|^2}
 {\veee{x}^{d+4}}\qquad[|y|\le\tfrac13|x|].\lbeq{D*n2}
\end{gather}
Applying those bounds to the analysis in \cite{cs14} for the random-walk's
Green function, we can choose the value of $\tK$ in \refeq{tKdef}
independently of $\lambda,\mu,N$.

Now, we prove the following theorem.
\begin{thm}\label{thm:Gbd}
Let $d>4$,
$N>\frac{2\hat\sJ^2}\lambda\vee\big(\frac{2\mu^2}\lambda\big)^3$ and
$\lambda\ll2\mu^2$.  Then, for any $\mu\in(\mu_N,\muNG]$,
\begin{align}
\bar G_\mu\equiv\sup_x\veee{x}^{d-2}G_N(x)\le2\tK.
\end{align}
\end{thm}

\Proof{Proof.}
First, we note that $\bar G_{\muNG}\le\tK$, due to
Proposition~\ref{prp:Gproperties}(i).  In order to complete the proof,
it thus suffices to show
\begin{enumerate}[(I)]
\item
continuity of $\bar G_\mu$ in $\mu\in(\mu_N,\muNG]$,
\item
existence of a forbidden region:
$\bar G_\mu\notin(2\tK,3\tK]$ for every $\mu\in(\mu_N,\muNG)$.
\end{enumerate}

In order to prove (I), we use the following lemma, which is a simple
adaptation of \cite[Lemma~5.13]{s06} to the current setting.

\begin{lmm}\label{lmm:equicont}
Let $\{g_\mu(x)\}_{x\in\Zd}$ be an equicontinuous family of functions in
$\mu\in[m,M]$, i.e., for any $\vep>0$, there exists a $\delta>0$ such
that $|\mu-\mu'|<\delta\Rightarrow|g_\mu(x)-g_{\mu'}(x)|<\vep$, uniformly
in $\mu,\mu'\in[m,M]$ and $x\in\Zd$.  If $\bar g_\mu\equiv\sup_xg_\mu(x)$ is
finite for each $\mu\in[m,M]$, then $\bar g_\mu$ is also continuous in
$\mu\in[m,M]$.
\end{lmm}

In order to apply this lemma to the current setting and prove continuity of
$\bar G_\mu$, it suffices to show that
$\{\veee{x}^{d-2}G_N(x)\}_{x\in\Zd}$ is an equicontinuous family
of functions in $\mu\in[m,\muNG]$, for every $m\in(\mu_N,\muNG)$.  However,
since $G_N(x)$ is an increasing limit of $G_{\tilde\Lambda_N}(o,x)$ as
$\Lambda\uparrow\Zd$, it boils down to show that $\veee{x}^{d-2}
\frac{\text{d}}{\text{d}\mu}G_{\tilde\Lambda_N}(o,x)$ is bounded uniformly in
$x\in\Zd$, $\Lambda\subset\Zd$ and $\mu\in[m,\muNG]$.  First, by using
Lebowitz' inequality~\cite{l74},
\begin{align}
&\bigg|\frac{\text{d}}{\text{d}\mu}G_{\tilde\Lambda_N}(o,x)\bigg|=\epsilon_N^2
 \bigg|\frac{\text{d}}{\text{d}I}\bigg(\frac{1-(N-1)\tanh I}N\EXP{\wsigma_o
 \wsigma_x}_{\tilde\Lambda_N}\bigg)\bigg|\nn\\
&\quad=\epsilon_N^2\bigg|-\frac{N-1}{N\cosh^2I}\EXP{\wsigma_o\wsigma_x}_{\tilde
 \Lambda_N}+\frac{1-(N-1)\tanh I}N\sum_{y\in\Lambda}\sum_{1\le i<j\le N}\EXP{
 \wsigma_o\wsigma_x;\sigma_{(y,i)}\sigma_{(y,j)}}_{\tilde\Lambda_N}\bigg|\nn\\
&\quad\le\epsilon_N^2\bigg(\frac{N-1}N\EXP{\wsigma_o\wsigma_x}_{\tilde
 \Lambda_N}+\frac{1-(N-1)\tanh I}N\sum_{y\in\Lambda}\sum_{\substack{i,j\in[N]\\
 (i\ne j)}}\EXP{\wsigma_o\sigma_{(y,i)}}_{\tilde\Lambda_N}\EXP{\sigma_{(y,j)}
 \wsigma_x}_{\tilde\Lambda_N}\bigg)\nn\\
&\quad\le\frac{\epsilon_N^2N}{1-(N-1)\tanh I}\bigg(G_{\tilde\Lambda_N}(o,x)
 +\sum_{y\in\Lambda}G_{\tilde\Lambda_N}(o,y)\,G_{\tilde\Lambda_N}(y,x)\bigg)
 \nn\\
&\quad\le\frac1\mu\bigg(G_N(x)+\sum_{y\in\Zd}G_N(y)\,G_N(x-y)\bigg).
\end{align}
By Proposition~\ref{prp:Gproperties}(ii), we have
$G_N(y-x)\le\bar K/\veee{y-x}^{d+\alpha}$, where
$\bar K=\max_{\mu\in[m,\muNG]}K_\mu$ and $\alpha$ is an arbitrarily large
number in the current setting.  Therefore, we arrive at
\begin{align}
\veee{x}^{d-2}\bigg|\frac{\text{d}}{\text{d}\mu}G_{\tilde\Lambda_N}(o,x)\bigg|
 &\le\frac{\bar K}{m\veee{x}^{2+\alpha}}+\frac{\veee{x}^{d-2}}m\sum_{y\in\Zd}
 \frac{\bar K}{\veee{y}^{d+\alpha}}\frac{\bar K}{\veee{y-x}^{d+\alpha}}
 \le\frac{C}{\veee{x}^{2+\alpha}},
\end{align}
where the constant $C$ is independent of $\Lambda$ and $\mu$.  This completes
the proof of (I).
\QED

Next, we prove (II) by showing that $\bar G_\mu\le3\tK$ implies
$\bar G_\mu\le2\tK$ for each $\mu\in(\mu_N,\muNG)$.  First, we derive an
identity for $G_N$ using \refeq{laceexp}.  Under the assumption
$\bar G_\mu\le3\tK$ and the hypothesis of the theorem,
we can use Proposition~\ref{prp:pibds} to obtain the $T\to\infty$ limit of
\refeq{laceexp}:
\begin{align}\lbeq{laceexp-lim}
\EXP{\sigma_{\tilde o}\sigma_{\tilde x}}_{\tilde\Lambda_N}=\pi_{\tilde
 \Lambda_N}(\tilde o,\tilde x)+\sum_{\tilde u,\tilde v\in\tilde\Lambda_N}
 \pi_{\tilde\Lambda_N}(\tilde o,\tilde u)\,(\tanh\tilde J_{\tilde u,\tilde
 v})\,\EXP{\sigma_{\tilde v}\sigma_{\tilde x}}_{\tilde\Lambda_N},
\end{align}
where $\pi_{\tilde\Lambda_N}(\tilde o,\tilde x)$ satisfies \refeq{pisumbd}.
Let
\begin{align}
\varPi_{\tilde\Lambda_N}(o,x)=\sum_{i,j\in[N]}\pi_{\tilde\Lambda_N}((o,i),(x,
 j)).
\end{align}
which satisfies
\begin{gather}
\bigg|\frac{\varPi_{\tilde\Lambda_N}(o,x)}N-\delta_{o,x}\bigg(1-\sum_{\tilde v}
 (\tanh\tilde J_{\tilde o,\tilde v})\EXP{\sigma_{\tilde v}\sigma_{\tilde o}}_{
 \tilde\Lambda_N}\bigg)\bigg|\le O(\tlamb^2)\bigg(\delta_{o,x}+\frac{O(\tlamb
 N)}{\veee{x}^{3(d-2)}}\bigg).
\end{gather}
Then, there is a subsequential limit
$\varPi_N\equiv\lim_{\Lambda_j\uparrow\Zd}\varPi_{\Lambda_j\times[N]}$ such
that, for every $x\in\Zd$,
\begin{align}\lbeq{varPibd}
\bigg|\frac{\varPi_N(x)}N-\delta_{o,x}\Big(1-\olzd\Big)\bigg|\le O(\tlamb^2)
 \bigg(\delta_{o,x}+\frac{O(\tlamb N)}{\veee{x}^{3(d-2)}}\bigg)
 \stackrel{\tlamb<\tlamb^2N}\le\frac{O(\tlamb^3N)}{\veee{x}^{3(d-2)}},
\end{align}
where
\begin{align}
\olzd=\sum_{\tilde v\in\tilde\mZ_N^d}(\tanh\tilde J_{\tilde o,\tilde v})
 \EXP{\sigma_{\tilde v}\sigma_{\tilde o}}_{\tilde\mZ_N^d}
 \stackrel{\refeq{bound-by-G2}}=O(\tlamb).
\end{align}
Therefore, the limit of the sum of \refeq{laceexp-lim} equals
\begin{align}
\EXP{\wsigma_o\wsigma_x}_{\tilde\mZ^d_N}
&=\varPi_N(x)+\sum_{\tilde u,\tilde v\in\tilde\mZ^d_N}\frac{\varPi_N(u)}
 N\,(\tanh\tilde J_{\tilde u,\tilde v})\,\frac{\EXP{\sigma_{\tilde v}
 \wsigma_x}_{\tilde\mZ^d_N}}N\nn\\
&=\varPi_N(x)+\sum_{u,v\in\Zd}\varPi_N(u)\,(\tanh J_{u,v})\,\EXP{\wsigma_v
 \wsigma_x}_{\tilde\mZ^d_N}\nn\\
&\quad+(N-1)(\tanh I)\sum_{v\in\Zd}\bigg(\frac{\varPi_N(v)}N-\delta_{o,v}\bigg)
 \EXP{\wsigma_v\wsigma_x}_{\tilde\mZ^d_N}\nn\\
&\quad+(N-1)(\tanh I)\,\EXP{\wsigma_o\wsigma_x}_{\tilde\mZ^d_N}.
\end{align}
Solving this identity for $\EXP{\wsigma_o\wsigma_x}_{\tilde\mZ^d_N}$ and
then dividing both sides by $N$, we obtain
\begin{align}\lbeq{laceexp-blockspin}
G_N(x)&=\frac{\varPi_N(x)}N+\sum_{u,v\in\Zd}\frac{\varPi_N(u)}N\,pD(v-u)\,G_N
 (x-v)\nn\\
&\quad+\frac{(N-1)\tanh I}{1-(N-1)\tanh I}\sum_{v\in\Zd}\bigg(\frac{\varPi_N
 (v)}N-\delta_{o,v}\bigg)G_N(x-v)\nn\\
&=\frac{\varPi_N(x)}N+\sum_{v\in\Zd}F_N(v)\,G_N(x-v),
\end{align}
where
\begin{align}\lbeq{Fdef}
F_N(x)=\sum_{u\in\Zd}\frac{\varPi_N(u)}N\,pD(x-u)+\frac{(N-1)\tanh I}{1-(N-1)
 \tanh I}\bigg(\frac{\varPi_N(x)}N-\delta_{o,x}\bigg).
\end{align}

Here, we note that, by summing \refeq{laceexp-blockspin} over $x\in\Zd$,
\begin{align}\lbeq{chiNdef}
\sum_{x\in\Zd}G_N(x)=\hat G_N(0)=\frac{\hat\varPi_N(0)/N}{1-\hat F_N(0)}.
\end{align}
Since $\hat\varPi_N(0)/N>0$ when $\tlamb^2N\ll1$ (cf., \refeq{varPibd}),
it must be that $\hat F_N(0)<1$ for $\mu>\mu_N$, which is equivalent to
\begin{align}\lbeq{muprebd}
\frac{\hat\varPi_N(0)}NN\sum_{x\in\Zd}\tanh J_{o,x}+(N-1)(\tanh I)
 \bigg(\frac{\hat\varPi_N(0)}N-1\bigg)&<1-(N-1)\tanh I\nn\\
\Leftrightarrow\quad N\sum_{x\in\Zd}\tanh J_{o,x}+(N-1)\tanh I
 &<\frac1{\hat\varPi_N(0)/N}.
\end{align}
Since $\sup_xJ_{o,x}<N^{-1}$ ($\because N>2\hat\sJ^2/\lambda$) and
$I=N^{-1}(1-o(1))$ ($\because N>(2\mu^2/\lambda)^3$),
the left-hand side of the above inequality is bounded below by
\begin{align}
N\sum_xJ_{o,x}\bigg(1-\frac{J_{o,x}^2}3\bigg)+NI\bigg(1-\frac{I^2}3\bigg)
 &\ge\big(1-O(N^{-2})\big)\bigg(N\sum_xJ_{o,x}+NI\bigg)\nn\\
&=\big(1-O(N^{-2})\big)\Big(1+(\hat\sJ-\mu)\epsilon_N^2N\Big).
\end{align}
As a result, \refeq{muprebd} implies
\begin{align}\lbeq{mubd}
&1+(\hat\sJ-\mu)\epsilon_N^2N<\frac{1+O(N^{-2})}{\hat\varPi_N(0)/N}\nn\\
&\Leftrightarrow\quad\mu>\hat\sJ-\frac{1+O(N^{-2})}{\epsilon_N^2N}\bigg(\frac1
 {\hat\varPi_N(0)/N}-1\bigg)-\frac{O(N^{-2})}{\epsilon_N^2N}\nn\\
&\quad\qquad=\hat\sJ-\underbrace{\frac{1+O(N^{-2})}{\epsilon_N^2N}\Big(\olzd
 +O(\tlamb^3N)\Big)}_{O(\lambda)/\mu}-\frac\lambda\mu\,O(N^{-4/3})=\hat\sJ
 -\frac{O(\lambda)}\mu,
\end{align}
where we have used $N>(2\mu^2/\lambda)^3$ to evaluate the $O(N^{-4/3})$ term.
Therefore, the assumed bound $\mu>\hat\sJ/2$ (cf., below \refeq{muNGbd}) is
indeed true if $\lambda\ll1$ (or $\hat\sJ\gg1$).

Next, we compare \refeq{laceexp-blockspin} with the convolution equation
for the random-walk's Green function:
\begin{align}\lbeq{RWconveq}
S_q(x)=\delta_{o,x}+(qD*S_q)(x).
\end{align}
Inspired by their similarity, we approximate $G_N$ by $r\frac{\varPi_N}N*S_q$,
with some $r\in(0,\infty)$ and $q\in[0,1]$.  In order to do so, we first
rearrange those convolution equations to get
\begin{align}
\frac{\varPi}N=G*(\delta-F),&&
\delta=(\delta-qD)*S_q,
\end{align}
where, for brevity, we have omitted the subscripts and the spatial variables.
Using those identities, we can rewrite $G$ as
\begin{align}\lbeq{G=varPi*S}
G&=r\frac{\varPi}N*S_q+G*\delta-r\frac{\varPi}N*S_q\nn\\
&=r\frac{\varPi}N*S_q+G*(\delta-qD)*S_q-rG*(\delta-F)*S_q\nn\\
&=r\frac{\varPi}N*S_q+G*E*S_q,
\end{align}
where
\begin{align}\lbeq{Edef}
E=(\delta-qD)-r(\delta-F).
\end{align}
We choose $q$ and $r$ to satisfy
\begin{align}\lbeq{qr-equations}
\begin{cases}
\hat E(0)=1-q-r\big(1-\hat F(0)\big)=0,\\[5pt]
\dpst\bar\nabla^2\hat E(0)\equiv\lim_{|k|\to0}\frac{\hat E(0)
 -\hat E(k)}{1-\hat D(k)}=-q+r\bar\nabla^2\hat F(0)=0,
\end{cases}
\end{align}
or equivalently
\begin{align}\lbeq{qr-equations2}
q=r\bar\nabla^2\hat F(0),&&
r=\frac1{1-\hat F(0)+\bar\nabla^2\hat F(0)}.
\end{align}
Then, we can rewrite $E$ as
\begin{align}
E=\delta-r\Big(\delta-F+\bar\nabla^2\hat F(0)D\Big)=r\Big(-\big(
 \hat F(0)\delta-F\big)+\bar\nabla^2\hat F(0)(\delta-D)\Big).
\end{align}
However, since (cf., \refeq{Fdef})
\begin{align}\lbeq{FFourier}
\hat F(k)=\frac{\hat\varPi(k)}Np\hat D(k)+\frac{(N-1)\tanh I}{1-(N-1)\tanh I}
 \bigg(\frac{\hat\varPi(k)}N-1\bigg),
\end{align}
we have
\begin{align}
\hat F(0)\delta-F=p\frac{\varPi}N*(\delta-D)+\bigg(p+\frac{(N-1)\tanh I}{1-(N-
 1)\tanh I}\bigg)\bigg(\frac{\hat\varPi(0)}N\delta-\frac{\varPi}N\bigg),
\end{align}
so that
\begin{align}\lbeq{nablaF}
\bar\nabla^2\hat F(0)=p\frac{\hat\varPi(0)}N+\bigg(p+\frac{(N-1)
 \tanh I}{1-(N-1)\tanh I}\bigg)\frac{\bar\nabla^2\hat\varPi(0)}N.
\end{align}
Therefore,
\begin{align}\lbeq{rrewr}
r=\bigg(1+\frac{(N-1)\tanh I}{1-(N-1)\tanh I}\bigg(1-\frac{\hat\varPi(0)}N+
 \frac{\bar\nabla^2\hat\varPi(0)}N\bigg)+p\frac{\bar\nabla^2\hat\varPi(0)}N
 \bigg)^{-1},
\end{align}
and
\begin{align}\lbeq{Erewr}
E&=r\Bigg(p\bigg(\frac{\hat\varPi(0)}N\delta-\frac{\varPi}N\bigg)*(\delta-D)
 \nn\\
&\qquad-\bigg(p+\frac{(N-1)\tanh I}{1-(N-1)\tanh I}\bigg)\bigg(\frac{\hat
 \varPi(0)}N\delta-\frac{\varPi}N-\frac{\bar\nabla^2\hat\varPi
 (0)}N(\delta-D)\bigg)\Bigg).
\end{align}
Due to those rewrites, we can show the following proposition,
whose proof follows after completion of the proof of (II).
\begin{prp}\label{prp:E*Sbd}
Let $q$ and $r$ be chosen to satisfy
\refeq{qr-equations}--\refeq{qr-equations2}.  Under the hypothesis of
Theorem~\ref{thm:Gbd}, there is a $\rho>0$ such that
\begin{align}\lbeq{E*Sbd}
r=1-O(\tlamb^2N),&&&&
0\le1-q\le O(\tlamb^2N),&&&&
|(E*S_q)(x)|\le\frac{O(\tlamb^2N)^2}{\veee{x}^{d+\rho}}.
\end{align}
\end{prp}

Finally, we can conclude $\bar G_\mu\le2\tK$ (hence (II)) by first rewriting
\refeq{G=varPi*S} as
\begin{align}
G=r\frac{\hat\varPi(0)}NS_q-r\bigg(\frac{\hat\varPi(0)}N\delta-\frac{\varPi}N
 \bigg)*S_q+G*E*S_q,
\end{align}
and then applying \refeq{varPibd}, Lemma~\ref{lmm:convbds} and
Proposition~\ref{prp:E*Sbd}.
This completes the proof of Theorem~\ref{thm:Gbd}.
\QED

\Proof{Proof of Proposition~\ref{prp:E*Sbd}.}
To evaluate $r$, we must investigate
$(N-1)(\tanh I)/(1-(N-1)\tanh I)$, $p$ and $\bar\nabla^2\hat\varPi(0)/N$ in
\refeq{rrewr}.  For the first two, it is easy to show that
\begin{align}
\frac{(N-1)\tanh I}{1-(N-1)\tanh I}\le\frac{NI}{1-NI}\le\frac1{\mu\epsilon^2
 N},
\end{align}
and that, by using $\hat\sJ<\mu+O(\lambda)/\mu$ (cf., \refeq{mubd}),
\begin{align}
p\le\frac{N\sum_xJ_{o,x}}{1-NI}=\frac{\hat\sJ}\mu=1+\frac{O(\lambda)}{\mu^2}
 =1+O(\tlamb^2N).
\end{align}
On the other hand, since $N>(2\mu^2/\lambda)^3$ (so that
$\mu\epsilon^2N<\tlamb^2N$), we have
\begin{align}
\frac{(N-1)\tanh I}{1-(N-1)\tanh I}&\ge\frac{(N-1)(I-\frac13I^3)}{1-(N-1)
 (I-\frac13I^3)}\ge\frac{(N-1)I-NI^3}{1-(N-1)I+NI^3}\nn\\
&\ge\frac{NI-\frac1N-\frac1{N^2}}{1-NI+\frac1N+\frac1{N^2}}\ge\frac{1-\mu
 \epsilon^2N(1+2\tlamb)}{\mu\epsilon^2N(1+2\tlamb)}\ge\frac{1-O(\tlamb^2N)}
 {\mu\epsilon^2N}.
\end{align}
Moreover, by using \refeq{muNGbd} and
$N>2(\hat\sJ\vee\mu)^2/\lambda$ (so that $\hat\sJ^2\epsilon^4<1/N^2$),
we have
\begin{align}
p\ge\frac{N\sum_x(J_{o,x}-\frac13J_{o,x}^3)}{1-(N-1)(I-\frac13I^3)}&\ge
 \frac{\hat\sJ\epsilon^2N(1-\hat\sJ^2\epsilon^4)}{\mu\epsilon^2N(1+2\tlamb)}
 \nn\\
&\ge\frac{N-1}N(1-\hat\sJ^2\epsilon^4)(1-2\tlamb)\ge1-O(\tlamb^2N).
\end{align}
Since $1/(\mu\epsilon^2N)=\tlamb N$, we can summarize the above bounds as
\begin{align}\lbeq{pest}
\frac{(N-1)\tanh I}{1-(N-1)\tanh I}=\tlamb N\big(1-O(\tlamb^2N)\big),&&
p=1+O(\tlamb^2N).
\end{align}
For $\bar\nabla^2\hat\varPi(0)/N$ in \refeq{rrewr}, we use
\refeq{varPibd} to obtain
\begin{align}\lbeq{hatvarPidiff}
\bigg|\frac{\hat\varPi(0)}N-\frac{\hat\varPi(k)}N\bigg|\le\sum_x
 \big(1-\cos(k\cdot x)\big)\bigg|\frac{\varPi(x)}N\bigg|\le O(\tlamb^3N)
 \sum_{x\ne o}\frac{1-\cos(k\cdot x)}{|x|^{3(d-2)}}.
\end{align}
However, since $d>4$, there is a $0<\tau<2\wedge(2(d-4))$ such that
\begin{align}\lbeq{1-cosdiff}
\sum_{x\ne o}\frac{1-\cos(k\cdot x)}{|x|^{3(d-2)}}=\frac{|k|^2}{2d}\sum_{x\ne o}
 \frac1{|x|^{d+2(d-4)}}+O(|k|^{2+\tau}).
\end{align}
Using \refeq{1-hatDasy}, we obtain
\begin{align}\lbeq{nablahatvarPiasy}
\frac{\bar\nabla^2\hat\varPi(0)}N=\lim_{k\to0}\frac{\hat\varPi(0)/N-\hat\varPi
 (k)/N}{1-\hat D(k)}=\frac{\sum_x|x|^2\varPi(x)/N}{\sum_x|x|^2D(x)}=O(\tlamb^3
 N).
\end{align}
As a result,
\begin{align}\lbeq{rest}
r=\bigg(1+\tlamb N\big(1-O(\tlamb^2N)\big)\Big(\olzd +\;O(\tlamb^3N)\Big)
 +O(\tlamb^3N)\bigg)^{-1}=1-O(\tlamb^2N).
\end{align}

To evaluate $1-q$ is straightforward.  By \refeq{qr-equations2} and
\refeq{FFourier}, we obtain
\begin{align}\lbeq{qest}
1-q=\underbrace{r\big(1-\hat F(0)\big)}_{\ge0~(\text{cf.,\,\refeq{chiNdef}})}
 =r\Bigg(1-\frac{\hat\varPi(0)}Np-\frac{(N-1)\tanh I}{1-(N-1)\tanh I}\bigg(
 \frac{\hat\varPi(0)}N-1\bigg)\Bigg)\nn\\
=r\Bigg(1-p+\bigg(p+\frac{(N-1)\tanh I}{1-(N-1)\tanh I}\bigg)\Big(\olzd+\;
 O(\tlamb^3N)\Big)\Bigg)=O(\tlamb^2N).
\end{align}

Finally, we investigate $(E*S_q)(x)$.  First, for a given $T\in(0,\infty)$, we
split it into two as
\begin{align}\lbeq{E*Sidentity}
(E*S_q)(x)&=\int_{[-\pi,\pi]^d}\frac{\text{d}^dk}{(2\pi)^d}\,\hat E(k)\frac{
 e^{-ik\cdot x}}{1-q\hat D(k)}\nn\\
&=\int_0^\infty\text{d}t\int_{[-\pi,\pi]^d}\frac{\text{d}^dk}{(2\pi)^d}\,\hat
 E(k)e^{-t(1-q\hat D(k))-ik\cdot x}\equiv X_{\sss>T}+X_{\sss<T},
\end{align}
where
\begin{align}
X_{\sss>T}&=\int_T^\infty\text{d}t\int_{[-\pi,\pi]^d}\frac{\text{d}^dk}{(2
 \pi)^d}\,\hat E(k)e^{-t(1-q\hat D(k))-ik\cdot x},\lbeq{X>def}\\
X_{\sss<T}&=\int_0^T\text{d}t\int_{[-\pi,\pi]^d}\frac{\text{d}^dk}{(2\pi)^d}\,
 \hat E(k)e^{-t(1-q\hat D(k))-ik\cdot x}.\lbeq{X<def}
\end{align}
The value of $T$ is arbitrary for now, but it is to be determined shortly.

Next, we estimate $X_{\sss>T}$ by taking the Fourier transform of
\refeq{Erewr} as
\begin{align}\lbeq{EFourier}
\hat E(k)&=r\big(1-\hat D(k)\big)\Bigg(p\bigg(\frac{\hat\varPi(0)}N-\frac{\hat
 \varPi(k)}N\bigg)\nn\\
&\qquad-\bigg(p+\frac{(N-1)\tanh I}{1-(N-1)\tanh I}\bigg)\bigg(\frac{\hat\varPi
 (0)/N-\hat\varPi(k)/N}{1-\hat D(k)}-\frac{\bar\nabla^2\hat\varPi
 (0)}N\bigg)\Bigg).
\end{align}
By \refeq{1-hatDasy} and using \refeq{pest}--\refeq{nablahatvarPiasy} to
evaluate the expression in the biggest parentheses of \refeq{EFourier},
we obtain
\begin{align}
|\hat E(k)|\le O(\tlamb^2N)^2|k|^{2+\tau}.
\end{align}
Therefore, by substituting this to \refeq{X>def} and using \refeq{1-hatDasy}
and $q\ge1-O(\tlamb^2N)$ (cf., \refeq{qest}), we obtain
\begin{align}\lbeq{X>bd1}
|X_{\sss>T}|\le\int_T^\infty\text{d}t\int_{[-\pi,\pi]^d}\frac{\text{d}^dk}{(2
 \pi)^d}\,|\hat E(k)|e^{-tq(1-\hat D(k))}&=O(\tlamb^2N)^2\int_T^\infty\text{d}t
 ~t^{-1-\frac{d+\tau}2}\nn\\
&=O(\tlamb^2N)^2\,T^{-\frac{d+\tau}2}.
\end{align}
Let
\begin{align}\lbeq{rhoTdef}
\rho=\frac{2\tau}{d+2+\tau},&&
T=\veee{x}^{2-\rho}.
\end{align}
Then, we arrive at
\begin{align}\lbeq{X>bd2}
|X_{\sss>T}|\le\frac{O(\tlamb^2N)^2}{\veee{x}^{d+\rho}}.
\end{align}

Next, we estimate $X_{\sss<T}$ by first expanding $e^{tq\hat D(k)}$ as
\begin{align}\lbeq{X<identity}
X_{\sss<T}&=\int_0^T\text{d}t\,e^{-t}\sum_{n=0}^\infty\frac{(tq)^n}{n!}
 \int_{[-\pi,\pi]^d}\frac{\text{d}^dk}{(2\pi)^d}\,\hat E(k)\,\hat D(k)^n\,
 e^{-ik\cdot x}\nn\\
&=\int_0^T\text{d}t\,e^{-t}\sum_{n=0}^\infty\frac{(tq)^n}{n!}(E*D^{*n})(x).
\end{align}
Since \refeq{Erewr} can be rearranged as
\begin{align}
E=r\Bigg(&-p\bigg(\frac{\hat\varPi(0)}N\delta-\frac{\varPi}N\bigg)*D-\frac{(N
 -1)\tanh I}{1-(N-1)\tanh I}\bigg(\frac{\hat\varPi(0)}N\delta-\frac{\varPi}N
 \bigg)\nn\\
&+\bigg(p+\frac{(N-1)\tanh I}{1-(N-1)\tanh I}\bigg)\frac{\bar\nabla^2\hat\varPi
 (0)}N(\delta-D)\Bigg),
\end{align}
we have
\begin{align}
(E*D^{*n})(x)=r\Bigg(&-p\sum_{y\ne o}\frac{\varPi(y)}N\Big(D^{*(n+1)}(x)-D^{*
 (n+1)}(x-y)\Big)\nn\\
&-\frac{(N-1)\tanh I}{1-(N-1)\tanh I}\sum_{y\ne o}\frac{\varPi(y)}N\Big(D^{*n}
 (x)-D^{*n}(x-y)\Big)\nn\\
&+\bigg(p+\frac{(N-1)\tanh I}{1-(N-1)\tanh I}\bigg)\frac{\bar\nabla^2\hat\varPi
 (0)}N\nn\\
&\hskip7pc\times\sum_{y\ne o}D(y)\Big(D^{*n}(x)-D^{*n}(x-y)\Big)\Bigg).
\end{align}
Suppose that
\begin{align}
\bigg|\sum_{y\ne o}\frac{\varPi(y)}N\Big(D^{*n}(x)-D^{*n}(x-y)\Big)\bigg|
 &\le\frac{O(\tlamb^3N)}{\veee{x}^{d+2}}\bigg(\frac{n}{\veee{x}^2}+1\bigg),
 \lbeq{varPi*Dbd}\\
\bigg|\sum_{y\ne o}D(y)\Big(D^{*n}(x)-D^{*n}(x-y)\Big)\bigg|&\le\frac{O(1)}
 {\veee{x}^{d+2}}\bigg(\frac{n}{\veee{x}^2}+1\bigg),\lbeq{D*Dbd}
\end{align}
so that, by \refeq{pest} and \refeq{rest},
\begin{align}
|(E*D^{*n})(x)|\le\frac{O(\tlamb^2N)^2}{\veee{x}^{d+2}}\bigg(
 \frac{n}{\veee{x}^2}+1\bigg).
\end{align}
Then, by \refeq{rhoTdef} and \refeq{X<identity}, we obtain
\begin{align}\lbeq{X<bd}
|X_{\sss<T}|\le\frac{O(\tlamb^2N)^2\,T}{\veee{x}^{d+2}}\bigg(
 \frac{T}{\veee{x}^2}+1\bigg)=\frac{O(\tlamb^2N)^2}{\veee{x}^{d
 +\rho}}.
\end{align}
Combining this with \refeq{E*Sidentity} and \refeq{X>bd2}, we obtain the
desired bound on $(E*S_q)(x)$, as in \refeq{E*Sbd}.

Now, it remains to show \refeq{varPi*Dbd}--\refeq{D*Dbd}.  Since their proofs
are almost identical, we only show here \refeq{varPi*Dbd}.  First, we split
the sum into three as
\begin{align}
\sum_{y\ne o}\frac{\varPi(y)}N\Big(D^{*n}(x)-D^{*n}(x-y)\Big)=\varSigma_1
 +\varSigma_2+\varSigma_3,
\end{align}
where
\begin{align}
\varSigma_1&=\sum_{y:0<|y|\le\frac13|x|}\frac{\varPi(y)}N\Big(D^{*n}(x)-D^{*n}
 (x-y)\Big),\\
\varSigma_2&=\sum_{y:|x-y|\le\frac13|x|}\frac{\varPi(y)}N\Big(D^{*n}(x)-D^{*n}
 (x-y)\Big),\\
\varSigma_3&=\sum_{y:|y|\wedge|x-y|>\frac13|x|}\frac{\varPi(y)}N\Big(D^{*n}(x)
 -D^{*n}(x-y)\Big).
\end{align}
We estimate $\varSigma_2$, $\varSigma_3$ and $\varSigma_1$ in order, by using
\refeq{D*n1}--\refeq{D*n2} and \refeq{varPibd}.

For $\varSigma_2$, since $|y|\ge|x|-|x-y|\ge\frac23|x|$ and $3(d-2)>d+2$ for
$d>4$, we obtain
\begin{align}
|\varSigma_2|\le\frac{O(\tlamb^3N)}{\veee{x}^{3(d-2)}}\sum_{y:|x-y|\le\frac13
 |x|}\Big(D^{*n}(x)+D^{*n}(x-y)\Big)\le\frac{O(\tlamb^3N)}{\veee{x}^{d+2}}
 \bigg(\frac{n}{\veee{x}^2}+1\bigg).
\end{align}
For $\varSigma_3$, we bound both $D^{*n}(x)$ and $D^{*n}(x-y)$ by
$O(n)/\veee{x}^{d+2}$ and use $3(d-2)>d+2$ again to obtain
\begin{align}
|\varSigma_3|\le\frac{O(n)}{\veee{x}^{d+2}}\sum_{y:|y|>\frac13|x|}\bigg|
 \frac{\varPi(y)}N\bigg|\le\frac{O(\tlamb^3N)}{\veee{x}^{d+4}}n.
\end{align}
For $\varSigma_1$, we first use the $\Zd$-symmetry of $\varPi$ and then use
\refeq{D*n2} to obtain
\begin{align}
|\varSigma_1|&=\bigg|\sum_{y:0<|y|\le\frac13|x|}\frac{\varPi(y)}N\bigg(D^{*n}
 (x)-\frac{D^{*n}(x+y)+D^{*n}(x-y)}2\bigg)\bigg|\nn\\
&\le\frac{O(n)}{\veee{x}^{d+4}}\sum_{y\ne o}|y|^2\bigg|\frac{\varPi(y)}N\bigg|
 \le\frac{O(\tlamb^3N)}{\veee{x}^{d+4}}n\sum_{y\ne o}\frac1{|y|^{d+2(d-4)}}
 =\frac{O(\tlamb^3N)}{\veee{x}^{d+4}}n.
\end{align}
This completes the proof of \refeq{varPi*Dbd}, hence the proof of
Proposition~\ref{prp:E*Sbd}.
\QED

\subsection{The linear Schwinger-Dyson equation}\label{ss:LSD}
Finally, we derive the linear Schwinger-Dyson equation \refeq{SDeq} and
complete the proof of the main theorem.

In the previous subsection, we have proved that, if $d>4$, $\lambda$ is
sufficiently small and $N$ is sufficiently large, then there is a $c<\infty$,
which is independent of $\lambda,\mu,N$, such that
$G_N(x)\le c/\veee{x}^{d-2}$ holds for all $x\in\Zd$ and
$\mu>\mu_N$.  Then, by Proposition~\ref{prp:pibds}, we have
\refeq{varPibd} uniformly in $\mu>\mu_N$.  Therefore, by
\refeq{laceexp-blockspin}--\refeq{Fdef}, we obtain that, for $\mu>\mu_N$,
\begin{align}\lbeq{laceexp-reapp}
G_N=\frac{\varPi_N}N+\Bigg(\frac{\varPi_N}N*pD+\frac{(N-1)\tanh I}{1-(N-1)
 \tanh I}\bigg(\frac{\varPi_N}N-\delta\bigg)\Bigg)*G_N.
\end{align}
Let
\begin{align}\lbeq{varPhiNdef}
\varPhi_N(x)=-\epsilon_N^2N^2\bigg(\frac{\varPi_N(x)}N-\delta_{o,x}\bigg)
 \stackrel{\refeq{varPibd}}=\epsilon_N^2N^2\olzd\delta_{o,x}+\frac{O(\lambda
 /\mu^3)}{\veee{x}^{3(d-2)}},
\end{align}
so that we can rewrite \refeq{laceexp-reapp} as
\begin{align}\lbeq{laceexp-rewr}
G_N=\delta-\frac{\varPhi_N}{\epsilon_N^2N^2}+\Bigg(pD-\bigg(pD+\frac{(N-1)
 \tanh I}{1-(N-1)\tanh I}\delta\bigg)*\frac{\varPhi_N}{\epsilon_N^2N^2}\Bigg)
 *G_N.
\end{align}

Now, we consider the $N\uparrow\infty$ limit of \refeq{laceexp-rewr}.  First,
we claim that $\lim_{N\uparrow\infty}\mu_N=\muc$.  In order to see this, we
recall that
\begin{align}
G_N(x)=\frac{1-(N-1)\tanh I}{\epsilon_N^2N}\epsilon_N^2\EXP{\wsigma_o
 \wsigma_x}_{\tilde\mZ^d}\ge0.
\end{align}
Since $\epsilon_N^2\EXP{\wsigma_o\wsigma_x}_{\tilde\Lambda_N}$ tends as
$N\uparrow\infty$ to $\Exp{\vphi_o\vphi_x}_\Lambda$ that is bounded
above by $\Exp{\vphi_o^2}_{\Zd}$ uniformly in $x$ and $\Lambda$ (due to
monotonicity and the Schwarz inequality), we can change the order of the
limits to obtain
\begin{align}\lbeq{change}
\lim_{N\uparrow\infty}\epsilon_N^2\EXP{\wsigma_o\wsigma_x}_{\tilde\mZ^d}
 =\Exp{\vphi_o\vphi_x}_\mu,
\end{align}
hence
\begin{align}
\lim_{N\uparrow\infty}G_N(x)=\mu\Exp{\vphi_o\vphi_x}_\mu.
\end{align}
Because of the nonnegativity of $G_N$, the $N\uparrow\infty$ limit of
$\sum_xG_N(x)$ is finite if and only if $\chi_\mu$ is finite.
This implies $\lim_{N\uparrow\infty}\mu_N=\muc$.

Suppose that, for every $\mu>\muc$, there is a subsequential limit
$\varPhi_\mu\equiv\lim_{N_j\uparrow\infty}\varPhi_{N_j}$ and that it is
summable.  Since
\begin{align}
\frac1{\epsilon_N^2N^2}=\sqrt{\frac\lambda{2N}},&&
pD\underset{N\uparrow\infty}\to\frac{\sJ}\mu,&&
\frac{(N-1)\tanh I}{1-(N-1)\tanh I}\underset{N\uparrow\infty}\sim\frac1\mu
 \sqrt{\frac{\lambda N}2},
\end{align}
we can take the limit of \refeq{laceexp-rewr} along this subsequence
to obtain
\begin{align}
\mu\Exp{\vphi_o\vphi_x}_\mu=\delta_{o,x}+\sum_v\bigg(\sJ(v)
 -\frac\lambda2\varPhi_\mu(v)\bigg)\Exp{\vphi_v\vphi_x}_\mu,
\end{align}
which is equivalent to the linear Schwinger-Dyson equation \refeq{SDeq}.

In order to complete the proof, it remains to show existence and summability
of the assumed subsequential limit
$\varPhi_\mu\equiv\lim_{N_j\uparrow\infty}\varPhi_{N_j}$.  However,
since the last term in \refeq{varPhiNdef} is summable uniformly in $N$,
we only need to show existence of the limit of the first term in
\refeq{varPhiNdef}.  Notice that, by \refeq{block-symmetry} (see
\refeq{block-symmetry-appl} as well),
\begin{align}
\epsilon_N^2N^2\olzd&=\epsilon_N^2N^2\sum_{\tilde v}(\tanh\tilde J_{\tilde o,
 \tilde v})\frac{\frac1N\EXP{\wsigma_v\wsigma_o}_{\tilde\mZ^d}-\delta_{v,o}}
 {N-\delta_{v,o}}\nn\\
&=\epsilon_N^2N(\tanh I)\Big(\EXP{\wsigma_o^2}_{\tilde\mZ^d}-N\Big)+
 \epsilon_N^2N\sum_v(\tanh J_{o,v})\EXP{\wsigma_v\wsigma_o}_{\tilde\mZ^d}\nn\\
&=N(\tanh I)\bigg(\epsilon_N^2\EXP{\wsigma_o^2}_{\tilde\mZ^d}-\epsilon_N^2N
 +\sum_v\frac{\tanh J_{o,v}}{\tanh I}~\epsilon_N^2\EXP{\wsigma_v\wsigma_o}_{
 \tilde\mZ^d}\bigg).
\end{align}
Since
\begin{align}
N\tanh I=1-\mu\epsilon_N^2N\underset{N\uparrow\infty}\to1,&&&&
\epsilon_N^2N=\sqrt{\frac2{\lambda N}}\underset{N\uparrow\infty}\to0,&&&&
\frac{\tanh J_{o,v}}{\tanh I}\underset{N\uparrow\infty}\sim\epsilon_N^2N\sJ(v),
\end{align}
we obtain (cf., \refeq{change})
\begin{align}
\lim_{N\uparrow\infty}\epsilon_N^2N^2\olzd=\Exp{\vphi_o^2}_\mu.
\end{align}
This completes the proof of the main theorem.
\QED

\section*{Acknowledgements}
This work is supported by the JSPS Grant-in-Aid for Scientific Research~(C)
24540106.  I would like to thank Markus Heydenreich for providing me with
the comfortable working environment during my stay in Leiden, March 2014.
I would also like to thank Gordon Slade and the anonymous referee for their
valuable comments to the earlier version of this paper.


\begin{thebibliography}{99}
\bibitem{a82}M. Aizenman.
\newblock Geometric analysis of $\phi^4$ fields and Ising models.
\newblock \emph{Comm. Math. Phys.}~\textbf{86} (1982): 1--48.

\bibitem{bbs14}R. Bauerschmidt, D.C. Brydges and G. Slade.
\newblock Scaling limits and critical behaviour of the 4-dimensional
$n$-component $|\vphi^4|$ spin model.
\newblock To appear in \emph{J. Stat. Phys.}

\bibitem{bfs82}D. Brydges, J. Fr\"ohlich and T. Spencer.
\newblock The random walk representation of classical spin systems and
correlation inequalities.
\newblock \emph{Comm. Math. Phys.}~\textbf{83} (1982): 123--150.

\bibitem{bfs83}D. Brydges, J. Fr\"ohlich and A. Sokal.
\newblock The random-walk representation of classical spin systems and
correlation inequalities.  II. The skeleton inequalities.
\newblock \emph{Comm. Math. Phys.}~\textbf{91} (1983): 117--139.

\bibitem{bs85}D. Brydges and T. Spencer.
\newblock Self-avoiding walk in 5 or more dimensions.
\newblock \emph{Comm. Math. Phys.}~\textbf{97} (1985): 125--148.

\bibitem{cs14}L.-C. Chen and A. Sakai.
\newblock Critical two-point functions for long-range statistical-mechanical
models in high dimensions.
\newblock To appear in \emph{Ann. Probab.}

\bibitem{cd12}C. Cotar and J.-D. Deuschel.
\newblock Decay of covariances, uniqueness of ergodic component and scaling
limit for a class of $\nabla\phi$ systems with non-convex potential.
\newblock \emph{Ann. Inst. H. Poincar\'e Probab. Statist.}~\textbf{48} (2012):
609-908.

\bibitem{fmrs87}J. Feldman, J. Magnen, V. Rivasseau and R. S\'en\'eor.
\newblock Construction and Borel summability of infrared $\varPhi_4^4$ by a
phase space expansion.
\newblock \emph{Comm. Math. Phys.}~\textbf{109} (1987): 437--480.

\bibitem{ffs92}R. Fern\'andez, J. Fr\"ohlich and A.D. Sokal.
\newblock \emph{Random walks, critical phenomena, and triviality in quantum
field theory} (Springer, Berlin, 1992).

\bibitem{f82}J. Fr\"ohlich.
\newblock On the triviality of $\lambda\vphi_d^4$ theories and the approach
to the critical point in $d\sideset{_{\raisebox{-1pt}{\tiny(}}}
{_{\raisebox{-1pt}{\tiny)}}}{\mathop\ge}4$ dimensions.
\newblock \emph{Nucl. Phys. B}~\textbf{200} (1982): 281--296.

\bibitem{fils78}J. Fr\"ohlich, R. Israel, E.H. Lieb and B. Simon.
\newblock Phase transitions and reflection positivity.  I. General theory and
long range lattice models.
\newblock \emph{Comm. Math. Phys.}~\textbf{62} (1978): 1--34.

\bibitem{fss76}J. Fr\"ohlich, B. Simon and T. Spencer.
\newblock Infrared bounds, phase transitions and continuous symmetry breaking.
\newblock \emph{Comm. Math. Phys.}~\textbf{50} (1976): 79--95.

\bibitem{gk80}K. Gaw\c{e}dzki and A. Kupiainen.
\newblock A rigorous block spin approach to massless lattice theories.
\newblock \emph{Comm. Math. Phys.}~\textbf{77} (1980): 31--64.

\bibitem{gk83}K. Gaw\c{e}dzki and A. Kupiainen.
\newblock Renormalization group for a critical lattice model.
\newblock \emph{Comm. Math. Phys.}~\textbf{88} (1983): 77--94.

\bibitem{gk84}K. Gaw\c{e}dzki and A. Kupiainen.
\newblock Asymptotic freedom beyond perturbation theory.
\newblock \emph{Critical Phenomena, Random Systems, Gauge Theories}
(Les Houches Summer School Proceedings, K. Osterwalder and R. Stora eds.,
North-Holland Physics Publishing, Amsterdam, 1984): 185--292.

\bibitem{gk85}K. Gaw\c{e}dzki and A. Kupiainen.
\newblock Massless lattice $\phi_4^4$ theory: Rigorous control of a
renormalizable asymptotically free model.
\newblock \emph{Comm. Math. Phys.}~\textbf{99} (1985): 197--252.

\bibitem{h08}T. Hara.
\newblock Decay of correlations in nearest-neighbour self-avoiding walk,
percolation, lattice trees and animals.
\newblock \emph{Ann. Probab.}~\textbf{36} (2008): 530--593.

\bibitem{hhs03}T. Hara, R. van der Hofstad and G. Slade.
\newblock Critical two-point functions and the lace expansion for spread-out
high-dimensional percolation and related models.
\newblock \emph{Ann. Probab.}~\textbf{31} (2003): 349--408.

\bibitem{hs90}T. Hara and G. Slade.
\newblock On the upper critical dimension of lattice trees and lattice animals.
\newblock \emph{J. Stat. Phys.}~\textbf{59} (1990): 1469--1510.

\bibitem{hs90'}T. Hara and G. Slade.
\newblock Mean-field critical behaviour for percolation in high dimensions.
\newblock \emph{Comm. Math. Phys.}~\textbf{128} (1990): 333--391.

\bibitem{hhs08}M. Heydenreich, R. van der Hofstad and A. Sakai.
\newblock Mean-field behavior for long- and finite-range Ising model,
percolation and self-avoiding walk.
\newblock \emph{J. Stat. Phys.}~\textbf{132} (2008): 1001--1049.

\bibitem{l74}J. Lebowitz.
\newblock GHS and other inequalities.
\newblock \emph{Comm. Math. Phys.}~\textbf{35} (1974): 87--92.

\bibitem{ny93}B.G. Nguyen and W-S. Yang.
\newblock Triangle condition for oriented percolation in high dimensions.
\newblock \emph{Ann. Probab.}~\textbf{21} (1993): 1809--1844.

\bibitem{s01}A. Sakai.
\newblock Mean-field critical behavior for the contact process.
\newblock \emph{J. Stat. Phys.}~\textbf{104} (2001): 111--143.

\bibitem{s07}A. Sakai.
\newblock Lace expansion for the Ising model.
\newblock \emph{Comm. Math. Phys.}~\textbf{272} (2007): 283--344.

\bibitem{s08}A. Sakai.
\newblock Applications of the lace expansion to statistical-mchanical models.
\newblock \emph{Analysis and Stochastics of Growth Processes and Interface
Models} (P. M\"oters et al. eds., Oxford University Press, Oxford, 2008):
123--147.

\bibitem{sg73}B. Simon and R.B. Griffiths.
\newblock The $(\phi^4)_2$ field theory as a classical Ising model.
\newblock \emph{Comm. Math. Phys.}~\textbf{33} (1973): 145--164.

\bibitem{s06}G. Slade.
\newblock The lace expansion and its applications.
\newblock \emph{Lecture Notes in Math.} \textbf{1879} (2006).

\bibitem{s82}A.D. Sokal.
\newblock An alternate constructive approach to the $\vphi_3^4$ quantum
field theory, and a possible destructive approach to $\vphi_4^4$.
\newblock \emph{Ann. Inst. Henri Poincar\'e Phys. Th\'eorique}~\textbf{37}
(1982): 317--398.
\end{thebibliography}
\end{document}